\PassOptionsToPackage{table}{xcolor}
\documentclass[
    reprint, 
    aps,
    preprintnumbers,
    twocolumn,
    prb,
    floatfix,
    superscriptaddress
]{revtex4-2}

\usepackage[utf8]{inputenc}
\usepackage{booktabs}
\usepackage[multiple]{footmisc}
\usepackage{rotating}

\usepackage{amssymb}
\usepackage{amsbsy}
\usepackage{amsmath}
\usepackage{bm}
\usepackage{bbm}
\usepackage{xfrac}
\usepackage{mathtools}

\usepackage{tikz}
\usepackage[T1]{fontenc}
\usepackage{array}
\usepackage{multirow}
\usepackage{etoolbox}
\usepackage{graphics}
\usepackage{url}
\usepackage{hyperref}
\usepackage{cleveref}
\usepackage{placeins}
\usepackage[normalem]{ulem}
\usepackage[]{algorithm2e}

\usepackage{adjustbox}

\graphicspath{{mainplots/}} %This is where you have saved the images

\newcommand{\vect}[1]{\mathbf{#1}}

\newcommand{\1}[1]{\textcolor{blue}{#1}}

\newcommand{\software}[1]{\textsc{#1}}
\newcommand{\package}[1]{\texttt{#1}}

\usepackage{xparse}
\NewDocumentCommand{\codeinline}{m}{\texttt{\detokenize{#1}}}

\usepackage{xcolor}
\usepackage{listings}
\usepackage{caption}
\usepackage{makecell}

\definecolor{codebg}{gray}{0.95}
\definecolor{kw}{RGB}{0,75,150}
\definecolor{num}{RGB}{150,0,0}
\definecolor{cmt}{RGB}{100,100,100}
\definecolor{sh}{RGB}{0,100,0}

\lstdefinestyle{base}{
  basicstyle=\ttfamily\small,
  backgroundcolor=\color{codebg},
  breaklines=true,
  breakatwhitespace=true,
  columns=fullflexible,
  keepspaces=true,
  showstringspaces=false,
  linewidth=\columnwidth,
  xleftmargin=0pt,
  xrightmargin=0pt,
  postbreak=\mbox{\(\hookrightarrow\)\space},
}

\lstdefinestyle{inline}{
  basicstyle=\ttfamily,
  backgroundcolor=\color{gray!20},
  frame=none,
  breaklines=false,
  linewidth=\columnwidth,
  xleftmargin=0pt,
  xrightmargin=0pt,
  showstringspaces=false
}

\lstdefinelanguage{mdp}{
  sensitive=true,
  morekeywords={
    integrator, dt, nsteps, continuation, tcoupl, tc-grps, ref_t,
    gen_vel, gen_temp, gen_seed, constraints, constraint_algorithm,
    comm_mode, energygrps, cutoff-scheme, ewald-rtol, vdwtype,
    rlist, rcoulomb, rvdw, pbc, define
  },
  morecomment=[l]{;},
  morecomment=[l]{\#},
  morestring=[b]"
}
\lstdefinestyle{mdp}{
  style=base,
  language=mdp,
  keywordstyle=\color{kw}\bfseries,
  commentstyle=\color{cmt}\itshape,
  numberstyle=\tiny\color{gray},
  numbers=left,
  frame=none
}

\lstdefinelanguage{namd}{
  sensitive=true,
  morekeywords={
    structure, coordinates, outputName, firsttimestep, numsteps,
    timestep, temperature, seed, outputEnergies, outputTiming,
    binaryoutput, cutoff, switching, pairlistdist, PME, langevin,
    langevinTemp, langevinDamping, run, wrapAll
  },
  morecomment=[l]{;},
  morecomment=[l]{\#},
  morestring=[b]"
}
\lstdefinestyle{namd}{
  style=base,
  language=namd,
  keywordstyle=\color{kw}\bfseries,
  commentstyle=\color{cmt}\itshape,
  numberstyle=\tiny\color{gray},
  numbers=left,
  numbersep=6pt,
  frame=none
}

\lstdefinelanguage{lammps}{
  sensitive=true,
  morekeywords={
    units, atom_style, read_data, pair_style, pair_coeff, neighbor,
    neigh_modify, timestep, fix, run, minimize, velocity, group, create_box
  },
  morecomment=[l]{\#},
  morestring=[b]"
}
\lstdefinestyle{lammps}{
  style=base,
  language=lammps,
  keywordstyle=\color{kw}\bfseries,
  commentstyle=\color{cmt}\itshape,
  numbers=left,
  numberstyle=\tiny\color{gray},
  numbersep=6pt,
  frame=none
}

\lstdefinelanguage{bashish}{
  sensitive=true,
  morekeywords={bash,sh,cp,mv,grep,sed,awk,tar,gunzip,gzip,python},
  morecomment=[l]{\#},
  morestring=[b]"
}
\lstdefinestyle{bashstyle}{
  style=base,
  language=bashish,
  keywordstyle=\color{sh}\bfseries,
  commentstyle=\color{cmt}\itshape,
  numbers=none,
  frame=single,
  rulecolor=\color{gray!60},
  basicstyle=\ttfamily\small
}

\lstdefinelanguage{python}{
  sensitive=true,
  morekeywords={import,from,as,def,return,if,elif,else,for,while,in,try,except,class,with,pass,break,continue,print,True,False,None,and,or,not,is,lambda,yield,global,nonlocal},
  morekeywords=[2]{len,str,int,float,list,dict,tuple,set,range,enumerate,zip,open,close,read,write},
  morecomment=[l]{\#},
  morestring=[b]',
  morestring=[b]",
  morestring=[s]{'''}{'''},
  morestring=[s]{"""}{"""}
}
\lstdefinestyle{pythonstyle}{
  style=base,
  language=python,
  keywordstyle=\color{kw}\bfseries,
  keywordstyle=[2]\color{blue!80}\bfseries,
  commentstyle=\color{cmt}\itshape,
  stringstyle=\color{sh},
  numberstyle=\tiny\color{gray},
  numbers=left,
  numbersep=6pt,
  frame=none,
  literate=%
    {0}{{{\color{num}0}}}1
    {1}{{{\color{num}1}}}1
    {2}{{{\color{num}2}}}1
    {3}{{{\color{num}3}}}1
    {4}{{{\color{num}4}}}1
    {5}{{{\color{num}5}}}1
    {6}{{{\color{num}6}}}1
    {7}{{{\color{num}7}}}1
    {8}{{{\color{num}8}}}1
    {9}{{{\color{num}9}}}1
}

\usepackage{subfig}  % Required for \subfloat command used in \oversubcaption
\usepackage{xstring} % Required for \stripFigNum command that uses \StrLen and \StrGobbleLeft
\newenvironment{captivy}[1]{%SETUP
  \begin{tikzpicture}[every node/.style={inner sep=0}]
    \node[anchor=south west,inner sep=0] (image) at (0,0) {#1};
    \begin{scope}[x={(image.south east)},y={(image.north west)}]
}%
{%TEARDOWN
        \end{scope}%
  \pgfresetboundingbox
  \path[use as bounding box] (image.south west) rectangle (image.north east);
  \end{tikzpicture}%
}

\newcommand*{\oversubcaption}[3]{%
  \draw (#1) node[fill=white,inner sep=0pt, opacity=0.2, above, yscale=1.1, xscale=1.1] {\phantom{(a)#2}};
  \draw (#1) node[inner sep=0pt, above]{%
    \subfloat[#2\label{#3}]{\phantom{(a)}}
  };
}

\newcommand{\stripFigNum}[1]{%
  \StrLen{\arabic{figure}}[\figlen]% how many chars in “1”, “2”, … “10”, …
  \StrGobbleLeft{#1}{\figlen}% drop that many from the front of #1
}

\Crefname{figure}{Fig.}{Figs.}
\Crefname{equation}{Eq.}{Eqs.}
\Crefname{section}{Sec.}{Secs.}
\Crefname{table}{Tab.}{Tabs.}
\creflabelformat{equation}{#2\textup{#1}#3}

\crefrangelabelformat{figure}{#3#1#4-#5\stripFigNum{#2}#6}

\begin{document}

\title[]{Streaming Molecular Dynamics Simulation Data for On-the-fly Processing and Analysis}

\author{Amruthesh Thirumalaiswamy}
\thanks{These authors contributed equally to this work.}
\affiliation{School of Molecular Sciences, Arizona State University, Tempe, AZ 85287, U.S.A.}

\author{Lawson J. Woods}
\thanks{These authors contributed equally to this work.}
\altaffiliation{formerly at: Translational Genomics Institute, Phoenix, AZ 85004, U.S.A.}
\affiliation{Department of Physics, Arizona State University, Tempe, AZ 85287, U.S.A.}
\affiliation{School of Molecular Sciences, Arizona State University, Tempe, AZ 85287, U.S.A.}

\author{Heekun Cho}
\affiliation{School of Molecular Sciences, Arizona State University, Tempe, AZ 85287, U.S.A.}

\author{Hugo MacDermott-Opeskin}
\author{Jennifer Clark}
\author{Irfan Alibay}
\author{Yuxuan Zhuang}
\affiliation{MDAnalysis, NumFOCUS, Austin, TX, U.S.A.}

\author{Oliver Beckstein}
\email{obeckste@asu.edu}
\affiliation{Department of Physics, Arizona State University, Tempe, AZ 85287, U.S.A.}
\affiliation{Center for Biological Physics, Arizona State University, Tempe, AZ 85287, U.S.A.}

\author{Matthias Heyden}
\email{mheyden1@asu.edu}
\affiliation{School of Molecular Sciences, Arizona State University, Tempe, AZ 85287, U.S.A.}
\affiliation{Center for Biological Physics, Arizona State University, Tempe, AZ 85287, U.S.A.}

\date{\today}

\begin{abstract}
Only a small fraction of the data generated in state-of-the-art all-atom multi-microsecond molecular dynamics (MD) simulations is typically analyzed. 
With femtosecond integration steps, microsecond simulations generate billions of time steps with a complete set of atomic positions, velocities, and forces for all atoms, corresponding to petabytes of data and exceeding typical storage capacities. 
Consequently, only a fraction of the simulated time steps are usually written to a trajectory file for subsequent analysis, often at time intervals of 10-100~ps. 
Such a trajectory file allows for the analysis of ensemble averages and slow dynamics, but information on faster processes is lost.
These fast processes include intra- and inter-molecular vibrations, dynamics in non-glassy solvents, short-lived transition states, transport properties, {\em etc.}, which encode physical information and are related to fundamental macroscopic properties and experimental observables.
Here, we introduce a data streaming interface for MD simulations that provides easy access to all data generated during a running simulation.
Instead of writing data to a storage medium, our interface enables user-defined analysis routines that access live simulation data via streaming.
For this purpose, we build on existing implementations of the Interactive Molecular Dynamics (IMD) protocol and implement an enhanced protocol (termed IMD version 3 or `IMDv3') in three popular MD packages \software{GROMACS}, \software{NAMD}, and \software{LAMMPS}.
Our new Python package \package{imdclient} receives an IMDv3 data stream and makes it available for other applications.
To maximize usability, we added the capability to process streamed MD simulation data to the popular \package{MDAnalysis} software package.
We demonstrate increased simulation performance for streaming compared to simulations that write data at high frequency to a trajectory output file.
We include usage examples that illustrate live monitoring of custom variables during a running simulation, evaluation of velocity time correlation functions with fast fluctuations, and instantaneous analysis of currents through a membrane pore.
\end{abstract}

\maketitle

\section{Introduction}
\label{sec:intro}

Molecular dynamics (MD) simulations are a powerful tool for studying physical systems at atomic and molecular scales, wherein particle trajectories are determined using Newton’s laws of motion.\cite{MDreviewLindahl2008,MDreviewHollingsworth2018,MDsimulationFrenkel2023} 
This approach is widely applied in areas such as soft matter, biological sciences, materials science, and nanotechnology.
In particular, MD serves as a valuable method for investigating molecular interactions and simulating the behavior of proteins, enzymes, and other biomolecules.\cite{MDreviewWarshel2002,MDreviewAdcock2006,BioMDHuggins2019}
Simulating material characteristics and properties through MD simulations drives avenues like material discovery in materials science and nanotechnology \cite{MaterialsMDAxelrod2022}.
Other popular use cases can be found in the field of chemical engineering, {\em e.g.}, in particle simulations of fluid dynamics and chemical separation \cite{ChemEngMDMaginn2010,FluidMDKarmakar2023,MDsimulationFrenkel2023}.

Complex systems often exhibit phenomena across a wide range of time and length scales.
The integration time steps of MD simulations are typically on the order of femtoseconds to resolve fast vibrations (needed for energy conservation) and stable integration of the equations of motion.
Consequently, the number of integration time steps that can be computationally performed for a given system provides an upper limit for the time scales that can be studied. 
In particular, many physically and biologically relevant processes, such as protein folding and phase separation, occur in microseconds to milliseconds \cite{ProteinDynamicsHenzler2007, SimulationTimeBowman2016}, which corresponds to billions to trillions of integration time steps.
Today, with the continual progress of Moore's Law as well as advances in supercomputing and custom hardware such as Anton \cite{SupercomputingShaw2021}, it is now feasible to simulate large biomolecular systems in atomic detail over a large range of timescales using unbiased MD simulations --- in effect capturing both fast vibrations and slow processes such as folding or conformational transitions.
However, analyzing fast (femtoseconds to picoseconds) processes in simulations carried out over timescales of multiple microseconds requires at least temporary storage of atomic coordinates, velocities or forces for millions or even billions of time steps. 
As a consequence, the analysis of intra- and inter-molecular vibrations, solvent dynamics and hydrogen bond fluctuations \cite{SolvationHeyden2019} in long simulations comes at the cost of excessive or even technologically unfeasible disk space requirements.
At the same time, frequent file I/O associated with the generation of femtosecond to picosecond resolution trajectories can have a substantial negative impact on the run-time performance of highly optimized simulation codes. 
Thus, while simulation trajectories that combine high time resolution with long simulation times have the potential to provide a wealth of information, scalable analysis is a challenge for the massive amount of data produced \cite{ComputingFoster2017, ConvergenceBeckstein2018}.

To mitigate file I/O and disk space requirements, atomic coordinates can be stored in trajectory files at coarse time intervals (e.g., every 100~ps), at the cost of losing high-frequency information \cite{ConvergenceBeckstein2018}. 
This limits the ability to track fast dynamical processes such as molecular vibrations, solvent relaxations, or barrier crossing events.
As a result, one loses the ability to compute many insightful time correlation and dynamic scattering functions, or to analyze important transport properties \cite{THzHeyden2010a, THzHeyden2010b, FreqAnalysisSauer2023, SolvationPersson2017, HydrophobicBeckstein2003, PorePermeationBeckstein2004, WeightedEnsembleRusso2022}.
Further, while most rare events in nature take milliseconds or much longer to occur, the transitions themselves happen on femtosecond to picosecond timescales \cite{TPSReviewBolhuis2002}.
Thus, observing the mechanism and process of barrier-crossing associated with events like defect migration in materials science \cite{DefectmigrationBai2010}, seed nucleation in reaction kinetics \cite{NucleationWedekind2007}, ion transfer in channels \cite{IonconductionJensen2010}, and short conformational angle flips in biomolecules \cite{ProteinDynShaw2010} requires high-frequency access to simulation data.

An alternative strategy to address the breadth of timescales is to combine results from multiple short simulations written out frequently and long simulations written out less frequently, to assess properties that emerge on log-time intervals. This, however, necessitates re-running trajectory segments at different time resolutions to properly stitch together the results. 
Such a complex simulation protocol inefficiently uses both human time and computational time and is furthermore complicated by the fact that on high-performance computing platforms leveraging GPUs, simulation trajectories are typically not fully reproducible because of non-deterministic rounding errors upon merging data from a large number of parallel threads. 
These challenges also make it difficult to study fast and slow processes in identical trajectories as required to investigate fast but rare free energy barrier crossing events that provide key mechanistic insights into enzyme function \cite{EnzymeDynamicsSchwartz2009}.

To address this problem, we propose an alternative to traditional file I/O (input/output) and the \textit{a posteriori} analysis of simulation trajectory files, namely to directly extract data as a live stream from a running simulation.
This idea has been recognized in multiple fields of computational science and appeared under various labels such as \emph{in-situ} or \emph{in-transit} analysis \cite{InsituBennett2012, DataEncodingLakshminarasimhan2013, ComputingFoster2017}. 
Streaming coordinate data for \emph{visualization} from MD simulations has been available since at least 2001 \cite{IMDoriginalStone2001, MDsrvKampfrath2022}. 
On the other hand, \emph{analysis} of MD simulations via streaming  has remained at the proof-of-principle stage with a focus on using advanced streaming frameworks \cite{FlexibleAnalysisDreher2014, InsituAnalysisJohnston2017, ScalableAnalysisMalakar2017, FlinkAnalysisZanuz2018}.
However, these approaches have found very limited adoption in the MD community, possibly because of the technical difficulty to make necessary code changes in the MD software and the lack of robust, easily installable, and flexible software for processing streamed trajectory data.
We therefore aimed for a streaming implementation that could be easily included in existing code bases and leverages proven technology on the producer (MD package) and receiver (analysis) side. 
The Interactive Molecular Dynamics (IMD) protocol, invented by Stone and coworkers in 2001 \cite{IMDoriginalStone2001, IMDv2Grayson2013}, proved to be an ideal starting point for our efforts. 
IMD is a simple package-based bidirectional protocol running over TCP/IP that was developed for real-time visualization and manipulation (\textit{e.g.} with haptic devices) of molecular dynamics simulations \cite{IMDreviewLanrezac2024}.
Version 2 implementation of this protocol (IMDv2) has been available in major MD codes such as \software{GROMACS} \cite{GROMACSPall2020}, \software{NAMD} \cite{NAMDPhillips2020}, and \software{LAMMPS} \cite{LAMMPSThompson2022} for many years.
However, the original IMD protocol and related implementations like IMDv2, were designed primarily for visualization (\textit{e.g.}, with \software{VMD} \cite{HumphreyVMD1996}) and lack the essential information needed for quantitative analysis. 
In particular, IMDv2 data streams do not include simulation time, simulation unit cell information, velocities, or forces.

In this work, we introduce a generalized streaming interface for molecular dynamics simulations built on top of existing IMDv2 implementations. 
Our approach extends current IMD functionalities by introducing a modified protocol that we call the IMD version 3 or IMDv3 protocol.
Implementations of this enhanced IMDv3 protocol allow flexible, real-time data streaming from simulations performed by popular high-performance MD simulation codes such as \software{GROMACS}, \software{NAMD}, and \software{LAMMPS}. 
In contrast to previous implementations, IMDv3 enables users to specify exactly which information (\textit{e.g.}, coordinates, velocities, forces, box dimensions) should be streamed (and which atoms or subsystems it should apply to). 
This approach allows for on-the-fly analysis of fast processes and system-specific events, eliminating the need for large intermediate files and extensive post-processing. By bridging the gap between data generation and analysis, our framework supports more efficient, scalable, and information-rich simulations.
As part of our work, we implemented IMDv3-streaming capability in \software{GROMACS}, \software{NAMD}, and \software{LAMMPS}. 
We created an open-source Python package \texttt{imdclient} for receiving and processing IMDv3 protocol specific simulation data from a TCP/IP socket connection. 
Using \texttt{imdclient}, we enabled the \texttt{MDAnalysis} package \cite{MDAnalysisGowers2016}, a popular platform for the analysis of MD simulation trajectories, to accept IMDv3 data streams for on-the-fly analysis.
Our setup not only enables seamless access to simulation data across a breadth of timescales, but is also a simple way to monitor any running MD simulation when needed.

\section{Methods}
\label{sec:methods}

\subsection{Streaming using the Interactive Molecular Dynamics protocol}
\label{subsec:imd-streaming}

In data streaming, information is produced at a source, transferred in real time through a network or local connection to a receiver, and immediately processed at the endpoint. 
The idea of streaming for molecular simulations was first presented by \citet{IMDoriginalStone2001} via \emph{Interactive Molecular Dynamics} (IMD), a setup wherein a user can interact or exchange data with a live running simulation. 
Simulation data is streamed as it is generated and received by a client, usually a visualization software, that can then render and display received data.
In this two-way communication setup, the client may also send certain information back to the simulation program, for instance to inject forces via a haptic device and thus directly influence the simulation itself in real time. 
For data transfer, IMD uses a fast, simple and proven network architecture via a TCP/IP (Transmission Control Protocol/Internet Protocol) socket.
A socket in this context is the software endpoint that connects the simulation software to the underlying network hardware through the computer's network interface using a port address. 
On the other end, a receiver (client) connects via its own socket to the same port address to enable the exchange (sending/receiving) of information. 
This communication is managed by the TCP/IP software stack, which encompasses a set of rules and protocols for handling the data transfer. 
IMD uses a custom application-level protocol (IMD protocol) that runs on top of a TCP/IP socket and sets the rules for sending and receiving specific simulation data.
Data as prescribed by this IMD protocol is sent in the form of data or message packets (more details in \Cref{subsubsecSI:imd-packet-types} of the Supporting Information (SI)).
The sequence of such data packets defined by the application-level protocol exchanged between the simulation and a client over a connection is termed an IMD data stream. 
IMD connects a \emph{producer} (simulation engine) and a \emph{receiver} (user program) so that they can interact and exchange information with each other as the simulation is running. 
Implementations in MD engines based on the IMD protocol, ``IMDv1'' \cite{IMDoriginalStone2001} and ``IMDv2'' \cite{IMDv2Grayson2013}, enable sending positions and energies from the simulation engine and receiving force feedback from the client and are mainly intended for interactive visualization purposes using \software{VMD} \cite{HumphreyVMD1996, IMDv2Grayson2013}.

\subsection{IMDv3 Protocol}
\label{subsec:imdv3-protocol}

We developed a modified, enhanced protocol, which we term the IMDv3 protocol, and is an extension of the original IMD protocol by \citet{IMDoriginalStone2001}.
Like the original protocol, IMDv3 provides for the sending of coordinates and energies from the simulation engine. 
In addition, this new protocol defines information packet types enabling the producer to send data types like simulation time, box dimensions, velocities, and forces.
Each information or data type is sent in the form of a message packet defined by its header and body (see \Cref{subsubsecSI:imd-packet-types} 
in the Supporting Information for more details).
However, in contrast to the original protocol, IMDv3 enforces a fixed order in which these packets are sent while providing the flexibility to configure, at the beginning of each connection, the particular selection of information being sent. 
The full technical specification is provided in the Supporting Information in \Cref{subsecSI:imd-specification}
and as part of the online documentation of the \package{imdclient} package \cite{IMDv3Woods2025}.

\subsection{New IMDv3 implementations}

We implemented the IMDv3 protocol in $3$ popular molecular simulation software packages, \software{GROMACS} \cite{GROMACSPall2020}, \software{LAMMPS} \cite{LAMMPSThompson2022},  and \software{NAMD} \cite{NAMDPhillips2020} that already included robust IMDv2 capabilities.
We re-used existing IMDv2 implementation code as much as possible and added additional features as required for IMDv3.
IMDv3 implementations can be used in the same context (\textit{e.g.}, with GPUs or MPI) as the existing IMDv2 ones.
The user has to select the IMD version to be used for communication with the client; with IMDv3, additional configuration options become available.
\Cref{subsecSI:md-engine-implementation} of the Supporting Information contains further details of the implementation.

As a reference implementation for the receiver we developed the \texttt{imdclient} Python package that provides the necessary programming infrastructure and tools to receive IMDv3 data from a server connected via a socket. 
The package uses an application-level buffer on the client end to temporarily store information parsed from the socket to preserve data and account for discrepancies in the speed of sending and receiving data. Implementation details can be found in \Cref{subsubsec:imdclient-implementation}.
Our reference Python implementation allows other applications to directly build on the IMDv3 protocol and is available under the permissive MIT license.

\subsection{Producer-Receiver Setup and Workflow}

\begin{figure*}
  \centering
  \includegraphics[width=0.75 \linewidth]{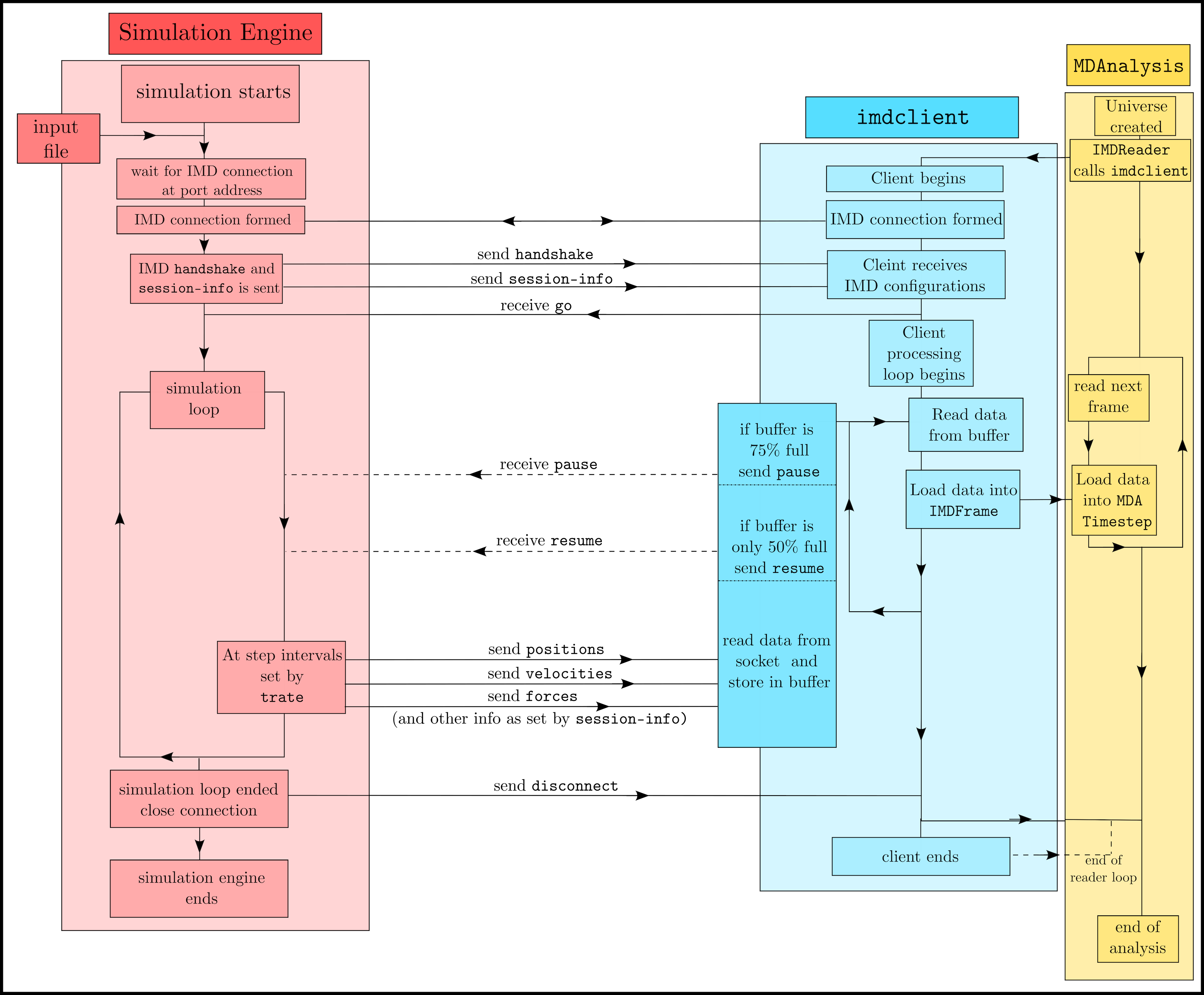}
  \caption{\textbf{Streaming Data using IMDv3:} The schematic illustrates the information flow when using IMDv3. The simulation engine (producer, red) interacts with the receiver (client, blue), namely the \package{imdclient} package, via a TCP/IP socket interface. Here, the \package{MDAnalysis} package (yellow) calls the client and loads the streaming data into standard MDAnalysis data structures. First, the client establishes a connection with the simulation engine, which can be configured to wait for such a connection prior to starting the simulation. Once the connection is established, the simulation engine starts the simulation and streams data to the client. \package{imdclient} processes this incoming data and stores data temporarily in an application-level buffer for eventual use by \package{MDAnalysis}.}
  \label{fig:imdv3-schematic}
\end{figure*}

The typical setup and flow of an IMDv3 session, as shown in the schematic \Cref{fig:imdv3-schematic}, includes the following steps:
\begin{enumerate}
  \item A simulation engine, configured with IMDv3 via command-line and input file settings, creates a socket and starts listening for connections on a user-defined port number such as 8888.
  \item A user program or package (for instance, \package{MDAnalysis} using the \package{imdclient} package), acts as IMDv3 receiver (client) and connects to the socket generated by the server, \textit{i.e.}, the simulation engine.
  \item Once the client and server are connected via the socket, the simulation engine sends a \codeinline{handshake} packet to \package{imdclient}, containing the endianness of the machine that it is running on together with the version of the IMD session (``3'' for IMDv3).
  \item The simulation engine sends a \codeinline{session-info} packet (introduced in IMDv3) to \package{imdclient} informing the client about the simulation data it should expect in this session, \textit{e.g.}, time, box dimensions, and positions.
  \item After parsing the \codeinline{handshake} and \codeinline{session-info} packets, \package{imdclient} sends a \codeinline{go} signal to the simulation engine and begins waiting for data packets.
  \item The simulation engine sends the specific data packets set by the user (in this example: time, box dimensions, and positions) in a fixed order for every $N$'th simulation frame, where $N$ is the user-defined IMD transmission rate or \codeinline{trate}.
  \item \package{imdclient} continuously parses the stream of bytes from the socket into its internal client buffer, making the data available as \package{numpy} arrays for processing.
  \item After ending its simulation loop, the simulation engine closes its socket and terminates the connection.
  \item \package{imdclient} recognizes that the session has ended and closes its socket connection after the last available frame of data has been processed into its internal buffer. \package{imdclient} exits after each IMD frame in the buffer has been processed.
\end{enumerate}

\subsection{Software availability and versions used}
\label{subsec:software-avail}
The source code for the \package{imdclient} package is available under the open source MIT license with the latest releases archived in Zenodo \cite{imdclientWoods2026}. The package can be installed from the Python Package Index with \texttt{pip install imdclient} or from the conda-forge channel with \texttt{mamba install -c conda-forge imdclient}. In this work we used imdclient release \codeinline{0.2.3}. 
\package{MDAnalysis} \cite{MDAnalysisGowers2016} supports reading IMDv3 streams since release \codeinline{2.10.0}. Its source code is available under the GNU Lesser General Public License, version \codeinline{2} or higher, releases are archived in Zenodo \cite{MDAGowers2025}, and it can be installed with \texttt{pip install MDAnalysis} or \texttt{mamba install -c conda-forge mdanalysis}. \package{MDAnalysis} will not automatically install \package{imdclient} so both packages need to be installed in order to process streams, \textit{e.g.}, \texttt{mamba install -c conda-forge mdanalysis imdclient}. Our benchmarking tests in \Cref{subsec:imd-benchmarking}, unless otherwise specified, were done using MDAnalysis version \codeinline{2.10.0}, archived in Zenodo \cite{MDAGowers2025}.

Support for IMDv3 in \software{GROMACS} \cite{GROMACSPall2020} has been implemented in a fork (see \Cref{tabSI:software-availability} of the Supporting Information
) of the \software{GROMACS} repository and is available under the same license as \software{GROMACS}, namely the GNU Lesser General Public License, version \codeinline{2.1}. This IMDv3 implementation used for our benchmarking later, was modified on top of \software{GROMACS} version \codeinline{2024.4}, and has been archived in Zenodo \cite{GMXIMDv3Woods2026}.

The IMDv3 implementation in \software{LAMMPS} \cite{LAMMPSThompson2022} has been a part of the official stable release since the \codeinline{22 July 2025} version. \software{LAMMPS} is available under the GNU General Public License, version \codeinline{2}. Additional optimizations to improve performance on GPUs when using streaming with the `kokkos'-mode \cite{kokkosLAMMPSJohansson2025} in \software{LAMMPS} are available in a fork (see \Cref{tabSI:software-availability} of the Supporting Information
) of the source code repository and archived in Zenodo \cite{LMPIMDv3Woods2026}. This particular version was used for our benchmarking efforts.

Support for IMDv3 in \software{NAMD} \cite{NAMDPhillips2020} has been merged into the official GitLab repository and will be publicly available in the upcoming 3.1 release. \software{NAMD} can be obtained from \url{https://www.ks.uiuc.edu/Research/namd/} under the University of Illinois \software{NAMD} Molecular Dynamics Software Non-Exclusive, Non-Commercial Use License. `GPU Resident mode' \cite{GPUresmodeNAMDPhillips2020} based optimizations for IMDv3 streaming are available on a branch of the source code repository currently, and were used for benchmarking in \Cref{subsubsec:lammps-namd-benchmarking}.

Links to the source code repositories, pre-IMDv3 commits/branches, IMDv3-enabled branches, particular versions used for benchmarking and their respective Zenodo DOIs are tabulated for reference in \Cref{tabSI:software-availability} of the Supporting Information.

Molecular images were prepared and rendered with VMD \cite{HumphreyVMD1996}.

\section{Results}
\label{sec:Results}

\subsection{New IMDv3 Capabilities in Simulation Engines}
\label{subsec:imd-implement-MD}

IMDv3 provides the ability to stream multiple types of simulation data and information. We have implemented this in \software{GROMACS}, \software{LAMMPS} and \software{NAMD} by building on existing IMDv2 implementations and enabling the user to switch between both versions.
To that effect, the user can use the following configuration settings in the simulation engine input file or execution command line.

\begin{itemize}
  \item Turning on IMD: Switch IMD on or off
  \item IMD port: The port address to which the socket connection is formed for streaming.
  \item Waiting behavior: Determines whether the simulation engine waits for incoming client connection before starting the simulation.
  \item Data transfer interval: Simulation step interval between communication of current frame information via IMD (historically also called transmission rate = \codeinline{trate}).
  \item IMD version: Setting the version and protocol type to be used
  \item Data types: Settings to configure which particular simulation data (box dimensions, positions, velocities, forces, etc.) should be streamed when using IMDv3. Some engines also offer the flexibility to stream a particular subset of the system (atoms).
\end{itemize}

\Cref{tab:imd-settings} summarizes the IMDv3 configuration parameters across all three simulation engines.

\begin{table*}[htbp]
\centering
\caption{IMDv3 configuration settings across \software{GROMACS}, \software{LAMMPS}, and \software{NAMD} simulation engines.}
\label{tab:imd-settings}
\renewcommand{\arraystretch}{1.3}
\begin{tabular}{p{4.5cm}p{3.5cm}p{3.5cm}p{3.5cm}}
\toprule
\multirow{2}{*}{\centering\textbf{IMD functionality}} & \multicolumn{3}{c}{\textbf{Simulation-specific setting}} \\
\cmidrule(lr){2-4}
& \textbf{\software{GROMACS}} & \textbf{\software{LAMMPS}} & \textbf{\software{NAMD}} \\
\midrule
Input filename & \codeinline{*.mdp} & \codeinline{*.in} & \codeinline{*.namd} \\
\midrule
Enable IMD & \codeinline{IMD-group} & \codeinline{fix imd} & \codeinline{IMDon} \\
Network port & \codeinline{-imdport}$^\dagger$ & \codeinline{port} & \codeinline{IMDport} \\
Protocol version & \codeinline{IMD-version} & \codeinline{version} & \codeinline{IMDversion} \\
Wait for initial client connection & \codeinline{-imdwait}$^\dagger$ & \codeinline{nowait} & \codeinline{IMDwait} \\
Data transfer rate & \codeinline{IMD-nst} & \codeinline{trate} & \codeinline{IMDfreq} \\
Atom group selection & \codeinline{IMD-group} & \codeinline{ID group-ID} & Not supported \\
\midrule
Stream simulation time & \codeinline{IMD-time} & \codeinline{time} & \codeinline{IMDsendTime} \\
Stream energy data & \codeinline{IMD-energies} & Not supported & \codeinline{IMDsendEnergies} \\
Stream box dimensions & \codeinline{IMD-box} & \codeinline{box} & \codeinline{IMDsendBoxDimensions} \\
Stream atomic positions & \codeinline{IMD-coords} & \codeinline{coordinates} & \codeinline{IMDsendPositions} \\
Stream atomic velocities & \codeinline{IMD-vels} & \codeinline{velocities} & \codeinline{IMDsendVelocities} \\
Stream atomic forces & \codeinline{IMD-forces} & \codeinline{forces} & \codeinline{IMDsendForces} \\
\midrule
Wrap/Unwrap across periodic boundaries & \codeinline{IMD-unwrap} & \codeinline{unwrap} & \codeinline{IMDwrapPositions} \\
\bottomrule
\end{tabular}\\[0.5em]
\raggedright
\footnotesize
$^\dagger$ Command-line flags specified at runtime, not in input file.
\end{table*}

It must be noted that even though the IMDv3 protocol supports force feedback, the current IMDv3 implementations in \software{LAMMPS} and \software{NAMD} do not support receiving and applying forces from a client.
The primary purpose and intention of developing IMDv3 was to enable easy access to simulation information that is relevant and suitable for scientific analysis.
Force feedback as a feature in IMDv2 was mostly intended for VMD-based interactive user sessions.
Thus, forces sent by the client at a certain timestep are not guaranteed to be received and applied at the exact same timestep on the simulation engine end.
The simulation engine receives and applies forces based on when and the order in which they are received, which in turn depends on IMD connection bandwidth speed and network latency.
Implementing synchronous force feedback would require significant changes to the protocol and would severely slow down the simulation.
Thus, our current IMDv3 implementation mainly focuses on expanding the ability to send analysis-relevant information to the client and improving compatibility with GPU-accelerated simulations.

In the following, we provide sample inputs for the three different simulation engines. More detailed information on these can be found in the Supporting Information.

\subsubsection{\software{GROMACS}}
\label{subsubsec:gromacs-imd-implement}
Example ( \codeinline{*.mdp}) input file snippet, to enable IMDv3 functionality in \software{GROMACS}:

\begin{lstlisting}[style=mdp]
IMD-group        = System
IMD-nst          = 100
IMD-version      = 3
IMD-time         = Yes
IMD-energies     = Yes
IMD-box          = Yes
IMD-coords       = Yes
IMD-vels         = Yes
IMD-forces       = Yes
IMD-unwrap       = Yes
\end{lstlisting}

\software{GROMACS} turns IMD functionality on/off using the \codeinline{IMD-group} input setting in the \codeinline{*.mdp} file. 
Further, \software{GROMACS} requires the user to define additional IMD-related variables, such as the port number and whether the simulation should wait for a client connection before starting, on the command line at runtime:

\begin{lstlisting}[style=bashstyle]
#!/bin/bash
gmx grompp -f input.mdp -c conf.gro -p topol.top -o run.tpr
gmx mdrun -v -deffnm run -imdport 8888 -imdwait
\end{lstlisting}

\subsubsection{\software{LAMMPS}}
\label{subsubsec:lammps-imd-implement}
Below is an example \software{LAMMPS} input file snippet to enable IMDv3 functionality:

\begin{lstlisting}[style=lammps]
fix imdv3 all imd 8888 nowait off trate 100 version 3 time yes box yes coordinates yes velocities yes forces yes unwrap yes 
\end{lstlisting}

\software{LAMMPS} can then be run as usual using the modified input file.

\subsubsection{\software{NAMD}}
\label{subsubsec:namd-imd-implement}
Example (\codeinline{*.namd}) input file snippet, to enable IMDv3 functionality in \software{NAMD}:

\begin{lstlisting}[style=namd]
IMDon                 yes
IMDport               8888
IMDwait               on
IMDfreq               100
IMDversion            3
IMDsendTime           yes
IMDsendEnergies       yes
IMDsendBoxDimensions  yes
IMDsendPositions      yes
IMDsendVelocities     yes
IMDsendForces         yes
IMDwrapPositions      no
\end{lstlisting}

Then, \software{NAMD} can be run as usual.

\subsection{IMDv3 based receivers}
\label{subsec:imd-implement-client}

Our new \package{imdclient} provides a pure Python reference implementation for an IMDv3 receiver.
Further, we used it to add the ability to accept IMDv3 streams instead of ordinary file-based trajectories to the \package{MDAnalysis} Python package \cite{MDAnalysisGowers2016}, thus enhancing \package{MDAnalysis}'s capabilities to seamlessly work with remote or cloud-based data sources \cite{ZarrtrajWoods2026}. 

\subsubsection{\package{imdclient}: a Python package for IMDv3 data processing}
\label{subsubsec:imdclient-implementation}

\begin{figure}[h!]
  \centering
  \includegraphics[width=\linewidth]{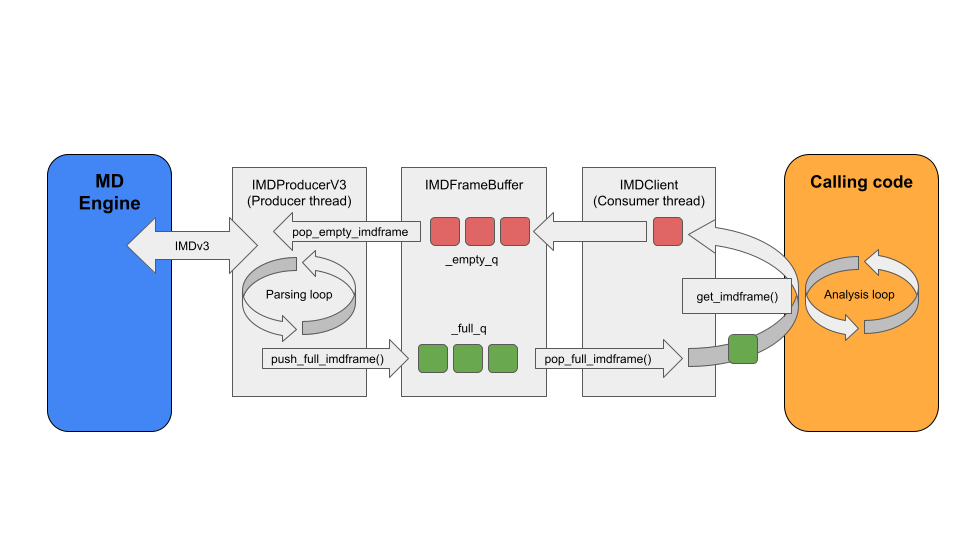}
  \caption{\textbf{Data processing workflow for \package{imdclient}:} Streaming data is received by \package{imdclient} via the TCP/IP interface and stored temporarily in an application-level buffer. This buffer is a fixed chunk of storage in the system's memory that operates as a FIFO (first-in-first-out) data structure implemented via two queues -- an empty queue which can store incoming data and a full queue which already has data that is ready to be processed. As the calling code requests newer frames, \package{imdclient} makes them available in the form of a Python dictionary from this full queue. 
  See \Cref{subsecSI:imdclient-implementation} of the Supporting Information 
  for more details.} 
  \label{fig:imdclient-arch}
\end{figure}

The \package{imdclient} package reads incoming streaming data following the IMDv3 protocol, processing it into an \codeinline{IMDFrame} object that contains all relevant simulation information, \textit{e.g.}, time, box dimensions, positions, velocities, forces, \textit{etc.}. The user is then free to use this object for further analysis. To avoid data loss and improve data transfer speeds between the client and the simulation engine, the package uses an application-level buffer to temporarily store simulation data information before processing them (\Cref{fig:imdclient-arch}). This fixed-memory buffer has been implemented in the form of two array queues. An empty queue contains memory elements that can store incoming data. As data is received and stored in the buffer elements, they move to the full queue. This efficient implementation allows us to avoid dynamic memory allocation and have a FIFO (first-in first-out) fixed memory buffer.
If the buffer approaches its capacity ($> 75\%$), for example, because the data is not processed (and removed from the buffer) quickly enough, \package{imdclient} sends a \codeinline{pause} packet to the simulation engine that temporarily pauses the simulation. Once the buffer is sufficiently emptied ($<50 \%)$, \package{imdclient} sends a \codeinline{resume} packet prompting the MD engine to resume the simulation. This implementation uses a multithreaded architecture to efficiently execute this functionality. 
\Cref{subsecSI:imdclient-implementation} in the Supporting Information provides further details.

Below, we provide a simple usage example where \package{imdclient} is used to connect to a running simulation engine at port number \codeinline{8888} to receive and process information from simulation frames.

\begin{lstlisting}[style=pythonstyle]
from imdclient.utils import parse_host_port
from imdclient.IMDClient import IMDClient 

# Simulation running on port number 8888
host, port = parse_host_port("imd://localhost:8888")

# Connect to the simulation.  n_atoms is needed 
# (e.g. extract from simulation input files).
# Explicit buffer size 100 MB (default 10 MB).
imdclient = IMDClient(host, port, n_atoms=n_atoms, buffer_size=100*1024*1024)

while True:
    try:
        imd_frame = imdclient.get_imdframe()
    except EOFError:
        break
    else:
        # Data available for further analysis in imd_frame
        print(imd_frame.time, imd_frame.energies, imd_frame.box, imd_frame.positions, imd_frame.velocities, imd_frame.forces)
\end{lstlisting}

The above example uses \package{imdclient} to establish a connection with a running simulation engine and receive/process data from it. 
The code then enters a loop to continuously access frames of data from the application-level buffer, making them accessible as a simple data structure. 
In practice, the user can write their custom routine that imports this data into other data structures of preference and perform on-the-fly analysis on it.

Further, the client package also provides the user with the option to toggle the simulation's waiting behavior at the beginning of an IMDv3 session using the \codeinline{continue\_after\_disconnect} argument while defining the \codeinline{IMDClient} object. 
This dictates whether the simulation waits for new connection or continues running when an existing IMD connection is detached.
Currently, \package{imdclient} doesn't support the ability to send forces to the simulation engine.
Full implementation details of the client package's features can be found in \Cref{subsecSI:imdclient-implementation} of the Supporting Information.

There are some inherent characteristics to the way data streams work and are handled by the \package{imdclient} package. Since data in the application-level buffer is automatically discarded after being processed by the client it can only be accessed in a sequential or forward-only manner. Unless data read from the stream is explicitly stored elsewhere (by the user), one cannot randomly jump to arbitrary frames in the stream or access previous frames. Thus, data in the stream is single-use and restart or rewind operations are not possible. 
Finally, the data in the stream is processed on-the-fly, and thus the total length of the data stream is unknown as the simulation runs. These properties are implied by the nature of data streams and any packages that use \package{imdclient} to stream simulation data thus inherit them.
If information from previous simulation frames is needed for the analysis, \textit{e.g.}, reference or lag-time separated states for time correlation functions, corresponding data structures (\textit{e.g.}, a ring buffer to store information on a specific number of past frames) and code to retain parts of the data stream need to be defined by the user.

\subsubsection{IMDv3-enabled \package{MDAnalysis}}
\label{subsubsec:mdanalysis-implementation}
The most evident use-case for IMDv3 streamed data is data analysis. To facilitate this application in a seamless manner, we implemented a corresponding \codeinline{Reader} class within \package{MDAnalysis} \cite{MDAnalysisGowers2016}, a popular Python package to read and analyze simulation data. The \codeinline{IMDReader} class enables the user to use \package{MDAnalysis}'s API to load and process IMDv3 streamed data using its port address just as one would for a data file on disk. 
\package{MDAnalysis} uses the \package{imdclient} package to process incoming streaming data and transfers per-frame information into its internal \codeinline{Timestep} data structures, ultimately making the streamed data accessible via its \codeinline{Universe} object. 

As an example we show how to use \package{MDAnalysis} to connect to a running simulation engine and calculate the end-to-end distance of a protein in the simulation for each time step in the simulation:
\begin{lstlisting}[style=pythonstyle]
import MDAnalysis as mda
from MDAnalysis.lib.distances import calc_bonds # for PBC-aware distance calculations

# Simulation running on port number 8888
# "topol.tpr" is the topology file for the system
u = mda.Universe("topol.tpr",
                 "imd://localhost:8888")
protein = u.select_atoms("protein")
for ts in u.trajectory:
    # Data available for analysis at each time step:
    # ts.time, ts.data["dt"], ts.data["step"], ts.data["energy_type"],
    # ts.dimensions
    # u.atoms.positions, u.atoms.velocities, u.atoms.forces

    # Using MDAnalysis methods for analysis:
    # Example distance calculation between protein's first and last atom
    distance = calc_bonds(protein[0].position, protein[-1].position, box=u.dimensions)
    print(f"Frame {ts.frame:4d} | Step: {ts.data['step']:4d} | Time: {ts.time:8.2f} ps | Distance: {distance:.2f} A")
\end{lstlisting}
Notably, the only indication that this code is processing a stream is the use of the URL-like string "imd://localhost:8888" as the ``trajectory'' filename for the \codeinline{Universe}.
The same code could be used to read a trajectory file from disk with only the line \codeinline{u = mda.Universe("topol.tpr", "trajectory.xtc")} changed.
Thus, the new streaming functionality extends the existing API seamlessly.
Our implementation of the \codeinline{IMDReader} class within \package{MDAnalysis} allows the user to treat a data stream just like a trajectory file on disk.

However, some of the technical limitations of streamed data directly impose restrictions on \codeinline{IMDReader} functionality in \package{MDAnalysis}.
Certain operations supported by other \codeinline{Reader} classes, \textit{e.g.}, slicing of a trajectory, random frame access, or global operations such as extracting data from all frames at once, are not supported by \codeinline{IMDReader}.
Some analysis classes within the \package{MDAnalysis.analysis} subpackage (MSD or mean-squared displacements, \codeinline{MDAnalysis.analysis.msd}; Hydrogen bond autocorrelation, \codeinline{MDAnalysis.analysis.hydrogenbonds.hbond_autocorrel}; {\em etc.}) require repeated or random access to frame data and are therefore currently not compatible with streamed trajectories.
These limitations can however be overcome by implementing user-defined routines and data structures ({\em e.g.}, ring buffers for storage) which temporarily store previously received frames for future analysis (as showcased in \Cref{subsec:imdv3-apps}).
On the other hand, analysis routines in \package{MDAnalysis} like radius of gyration, center of mass, root mean square deviation, distance calculation etc., which are based on data from a single frame, work without any modifications with the \codeinline{IMDReader} when called using low-level helper functions ({\em e.g.}, \codeinline{MDAnalysis.analysis.rms}, \codeinline{MDAnalysis.core.groups.center_of_mass()} etc.)
In addition to that, the current implementation of \codeinline{IMDReader} cannot be used with parallel implementations that would generate copies of the reader because only a single data stream is supported.
Lastly, writing data streams to a socket connection is currently not supported in \package{imdclient} and therefore no \codeinline{IMDWriter} is currently available in \package{MDAnalysis}.

\subsection{Benchmarking IMDv3 streaming}
\label{subsec:imd-benchmarking}
The use of IMD-streaming negates the need to store large data files when needing to access simulation data frequently. In the following, we investigate if and how the use of streaming slows down the simulation engine and under what streaming conditions it provides a performance advantage over file I/O.

We used a simulation of hen egg-white Lysozyme (HEWL) in water consisting of $\sim 30,000$ atoms as a test system for \software{GROMACS} and \software{NAMD}. For \software{LAMMPS}, we used a system of bead-spring polymers with finite extensible nonlinear elastic bonds containing $32,000$ beads. For each simulation engine, we analyzed the simulation performance (simulated time per wall-clock time) in different scenarios as discussed below.

Before comparing the performance of streaming and file I/O, we determined optimal computational settings for running the test system (with streaming and file I/O turned off), given a fixed set of computational resources, {\em i.e.}, one node with one A100 GPU and $48$ (2x AMD EPYC 7413 Zen3) cores (CPUs). All benchmarking simulations were run on Arizona State University's SOL supercomputer \cite{SOLJennewein2023}.
Furthermore, we verify that introducing our IMDv3 implementation does not impact the performance of the unmodified simulation engine.
The particular versions of \software{GROMACS}, \software{LAMMPS} and \software{NAMD} used for our benchmarking tests are discussed in \Cref{subsec:software-avail} with links available in \Cref{tabSI:software-availability} of the Supporting Information.
Benchmarking data, related analysis, and plotting functions are available on GitHub (see \Cref{tabSI:software-availability} of the Supporting Information) and archived on Zenodo \cite{IMDv3perfThirumalaiswamy2026}.

\subsubsection{GROMACS}
\label{subsubsec:gromacs-benchmarking}

We consider versions of \software{GROMACS} pre and post-IMDv3 modifications, which we call the `vanilla' and `imdv3' versions.
Running simulations with these two versions, varying the number of cores used, over a combination of thread-MPI tasks and OpenMP threads, tells us two important things. 
Evidently, under the same conditions the IMDv3-modified version of \software{GROMACS} consistently reaches the same performance as the unmodified version of \software{GROMACS}, indicating that the IMDv3 related code changes do not affect the performance of a \software{GROMACS} simulation.
Second, it gives us an optimum performance around $24$ cores using a single thread-MPI task and $24$ OpenMP threads (see \Cref{figSI:GROMACS-optimization} of the Supporting Information). This optimum performance serves as a reference for all our \software{GROMACS} benchmarks. %, as shown in \Cref{figSI:GROMACS-optimization}.

We compare and contrast IMDv3 streaming with file I/O, under different run settings. In particular, we use the streaming interval ($n_\text{s}$), {\em i.e.}, \codeinline{trate} or the file I/O interval ($n_\text{IO}$) as our parameter of interest to vary the run conditions. 
We consider two sets of streaming/output data types --- one with just positions (\codeinline{x}), and another with positions, velocities and forces (\codeinline{xvf}). 
For the former, we separately tested file I/O for writing positions to a compressed binary \codeinline{*.xtc} file and a full precision \codeinline{*.trr} file. 
For positions, velocities, and forces (not supported by \codeinline{*.xtc} files), we tested file I/O only for the \codeinline{*.trr} format.
File I/O is compared to streaming while streamed data is processed by \package{imdclient} without any analysis or output.
It is worth noting that any client-side analysis or output routine that processes and analyzes streamed data faster than the rate at which it is streamed by the simulation will not impact simulation performance.
However, slow processing of the streamed data will slow down the simulation (via \codeinline{pause-resume} messages from the client; see \Cref{subsubsec:imdclient-implementation} for details).
For comparison purposes, we restrict our analysis to simple client processing without any analysis/output.

\begin{figure}[h!]
  \centering
  \includegraphics[width=\linewidth]{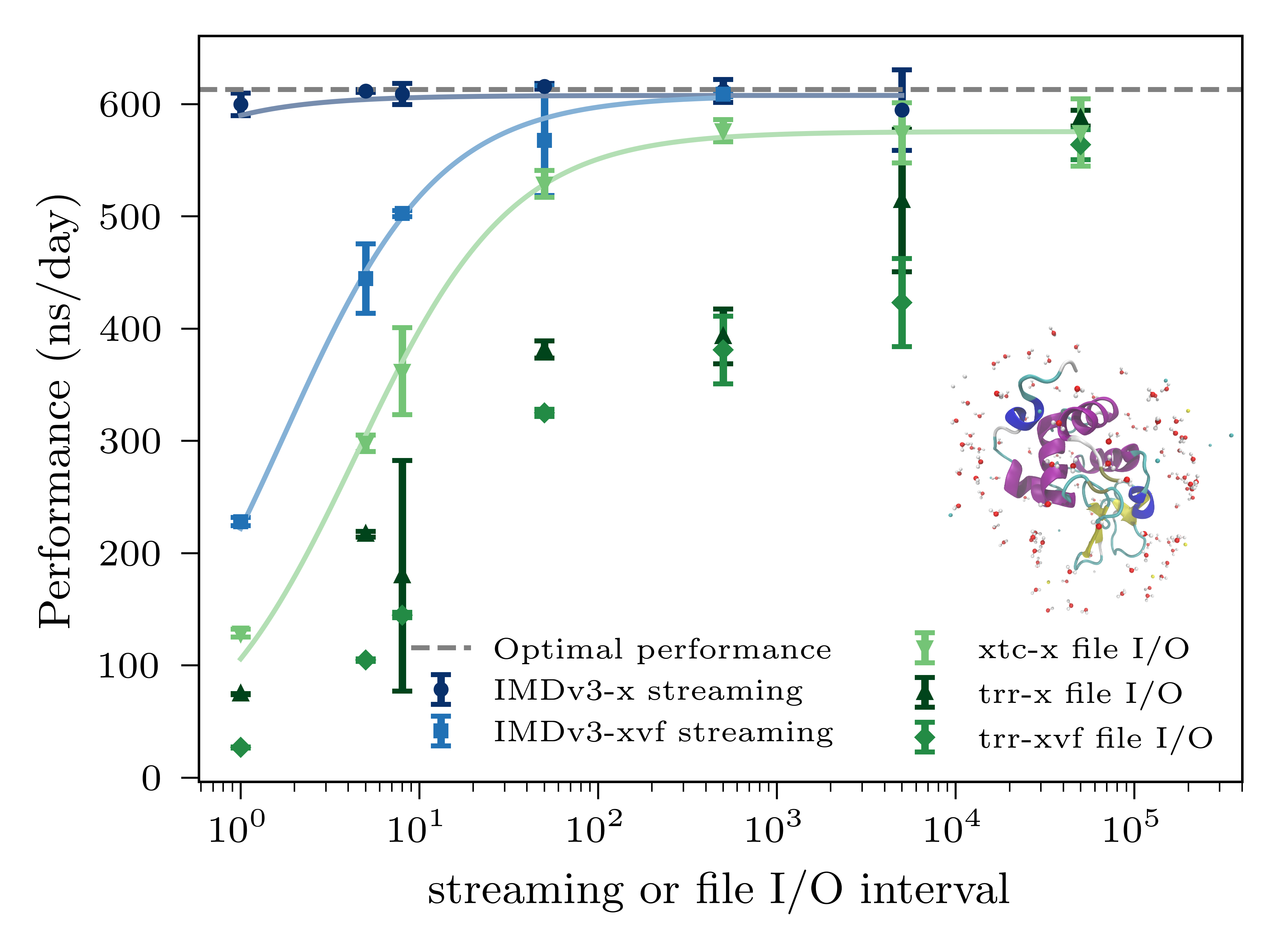}
  \caption{
  \textbf{Streaming is up to $\sim 5\times$ faster than file I/O in \software{GROMACS}:} We compare various streaming- and file I/O-based scenarios in \software{GROMACS} for the solvated hen egg-white lysozyme system (inset). For conventional data writing to files, we run three different scenarios compared, namely writing positions to a TRR (trr-x) or an XTC file (xtc-x) and writing positions-velocities-forces to a TRR file (trr-xvf). These are compared with streaming the same set of data (IMDv3-x and IMDv3-xvf) over to a receiver that simply processes and loops through incoming frames. Streaming provides a significant advantage, which is most pronounced at the shortest streaming/file I/O intervals. The gray dashed line at the top provides the optimal baseline performance achieved without any file I/O or streaming (see \Cref{figSI:GROMACS-optimization} of the Supporting Information). Solid curves resulted from fitting the entire \software{GROMACS} data set to the model described in \Cref{eq:master-sim-model}. This performance model successfully captures the behavior observed for both streaming and the xtc file I/O scenario. Curves for the TRR cases are not shown as the fits fail to model the eventual plateauing observed at high file I/O intervals.
  }
  \label{fig:GROMACS-benchmarking}
\end{figure}

Our benchmarking reveals the advantages that IMD streaming offers over traditional file I/O (see \Cref{fig:GROMACS-benchmarking} for \software{GROMACS}). We demonstrate that MD performance depends significantly on the file I/O interval. Writing or outputting files every time step or every few time steps significantly reduces the simulation performance for \software{GROMACS} -- slowing it down by $\sim 5$ times compared to the baseline performance without file I/O.
Simulation performance improves when data is written to an output file less frequently with an approximately log-linear relationship of performance and file I/O interval.
Using a compressed binary \codeinline{*.xtc} format, as opposed to a simple binary format like \codeinline{*.trr} for file I/O, improves performance slightly (by about $2$ times) and reaches the optimal performance of $\sim 600$ ns/day at shorter output intervals.

Streaming, on the other hand, is significantly faster -- streaming just positions (\codeinline{IMDv3-x}) is seemingly agnostic of the streaming interval and we observe a consistent simulation performance of $\sim 600$ ns/day even when streaming coordinates (\codeinline{IMDv3-x}) at every simulation step (streaming interval $= 1$). When velocities and forces are additionally streamed (\codeinline{IMDv3-xvf}), we see a similar trend as with file I/O -- performance improves with increasing data transfer interval before plateauing -- but continues to be faster than any of the file I/O scenarios. This provides clear evidence for the performance advantages of data streaming with IMDv3 compared to file I/O. 

At the same time, we note that even \codeinline{IMDv3-xvf} streaming at every simulation step only rarely triggers \codeinline{pause} messages ($\sim 1-2$ throughout the simulation) from \package{imdclient}, indicating that the performance reduction is not due to the ability of \package{imdclient} to accept data.
For a bottleneck in receiving or processing data with \package{imdclient}, we would expect frequent \codeinline{pause}-\codeinline{resume} messages.
Their absence indicates that the observed slowdown is caused by a bottleneck in the simulation engine, {\em e.g.}, due to gathering data from parallel processes for output or the engine failing to send data fast enough via the TCP/IP socket.

\subsubsection{Theoretical model for performance}
\label{subsubsec:model}

The asymptotic $1/\text{\codeinline{interval}}$ trend in computational performance observed for the \codeinline{IMDv3-xvf} case in \Cref{fig:GROMACS-benchmarking} can be explained using a simple modular model that accounts for the average time taken by the simulation engine to compute a time step ($t_{\text{MD}}$) and the time to then stream data generated at that time step ($t^{(\alpha)}_{\text{s}}$); this model was initially based on inspection of the code flow in \software{GROMACS} but has also shown itself to be applicable to the other simulation engines. 
The per-step time $t^{(\alpha)}_{\text{s}}$  may vary across simulation runs depending on the amount/type of data being streamed, with $\alpha$ denoting the various scenarios of streaming considered for, {\em e.g.}, $t^{(\text{x})}_{\text{s}}$ when streaming just positions and $t^{(\text{xvf})}_{\text{s}}$ when streaming positions, velocities, and forces.
The total wallclock time for the simulation is then calculated as the sum of simulation and streaming times. 
For a given data streaming interval, $t^{(\alpha)}_{\text{s}}$ only contributes every $n_\text{s}$ steps, while $t_{\text{MD}}$ contributes to every time step.
Additionally, at each time step the IMD module/implementation in the simulation engine checks for incoming message packets from the client, resulting in a constant per step cost ($t^{(0)}_{\text{s}}$).
Taken together, the total computational time for a simulation running for $N_\text{steps}$ timesteps is
\begin{equation*}
  {t_{\text{comp}}} = N_\text{steps} \times (t_{\text{MD}} + t^{(\alpha)}_\text{s}/n_\text{s} + t^{(0)}_{\text{s}}).
\label{eq:time-sim-model}
\end{equation*}
The simulation performance
\begin{equation}
  \mathfrak{p} = \frac{t_\text{sim}}{t_\text{comp}} = \frac{N_\text{steps} \times \Delta t}{t_\text{comp}}
    = \frac{\Delta t}{t_{\text{MD}} + \frac{t^{(\alpha)}_\text{s}}{n_\text{s}} + t^{(0)}_{\text{s}}}
\label{eq:speed-sim-model}
\end{equation}
is calculated as the ratio of simulation and computational time where $\Delta t$ is the simulation time step.
We note that file I/O simulations show a similar asymptotic-like trend. Thus, we generalize our model (\Cref{eq:speed-sim-model}) to account for various file I/O and streaming scenarios through a global model as follows,
\begin{equation}
  \mathfrak{p} = \frac{\Delta t}{t_{\text{MD}} + [\text{IMDon}]\left([n_\text{s}] \frac{t^{(\alpha)}_{\text{s}}}{n_\text{s}} + t^{(0)}_{\text{s}}\right) + [n_\text{IO}]\left(\frac{t^{(\alpha, \beta)}_{\text{IO}}}{n_\text{IO}} + t^{(\alpha, \beta, 0)}_{\text{IO}}\right)}
\label{eq:master-sim-model}
\end{equation}
where $[\text{IMDon}]$, $[n_\text{s}]$ and $[n_\text{IO}]$ are indicator functions that take the value of $1$ when IMD streaming is turned on or streaming/file I/O interval is $>0$, respectively, and $0$ otherwise.
The respective streaming and file I/O times denoted by $t^{(\alpha)}_\text{s}$ and $t^{(\alpha, \beta)}_\text{IO}$ vary based on the type and amount of information being streamed/output.
Specifically, $\alpha$ captures the type of data being streamed/output {\em i.e.}, $\alpha \in \{ \text{x}, \text{xvf} \}$ and $\beta$ represents the output format for file I/O with $\beta \in \{ \text{xtc}, \text{trr} \}$.
While the constant streaming cost $t^{(0)}_{\text{s}}$ is treated the same across all streaming scenarios, the file I/O constant cost $t^{(\alpha, \beta, 0)}_{\text{IO}}$ is treated as separate values for each different file I/O scenario.

The expression for simulation performance \Cref{eq:master-sim-model} (defined in, for example, ns/day), allows us to fit for $t_{\text{MD}}$, $t^{(\text{x})}_{\text{s}}$, $t^{(\text{xvf})}_{\text{s}}$ using the performance values from our reference baseline and various scenarios at distinct streaming intervals. 
One can then describe the expected benchmark performance for any data streaming intervals, using the expression $\Delta t /(t_{\text{MD}} + t^{(\alpha)}_\text{s}/n_\text{s} + t^{(0)}_\text{s})$, as shown in \Cref{fig:GROMACS-benchmarking}.
Apart from describing our benchmark results, these fits also reveal that the time taken by the simulation engine to stream data for the \codeinline{IMDv3-x} and \codeinline{IMDv3-xvf} cases are dramatically different with $t^{({\text{x}})}_{\text{s}} \ll t^{({\text{xvf}})}_{\text{s}}, t_{\text{MD}}$ (see \Cref{tabSI:fit-params-all-engines} of the Supporting Information).
For file I/O, while the model captures the scenario of writing positions to an XTC file well, it fails to do so for TRR files. We suspect that other code functions related to TRR may run in parallel and thus invalidate the serial modularity assumptions of our model.

\subsubsection{\software{LAMMPS} and \software{NAMD}}
\label{subsubsec:lammps-namd-benchmarking}

We perform similar benchmarks for \software{LAMMPS} and \software{NAMD} to compare performance between streaming and file I/O. 
For \software{NAMD}, we use the same biomolecular system (solvated HEWL) as described for \software{GROMACS}.
For \software{LAMMPS}, we choose a system that more closely resembles typical use cases in the corresponding community. Here, we use a simple FENE-bonded polymer system with $320$ polymers with $100$ monomers each, which results in approximately the same number of particles as the system used for \software{GROMACS} and \software{NAMD}.

\begin{figure}[h!]
  \centering
  \includegraphics[width=\linewidth]{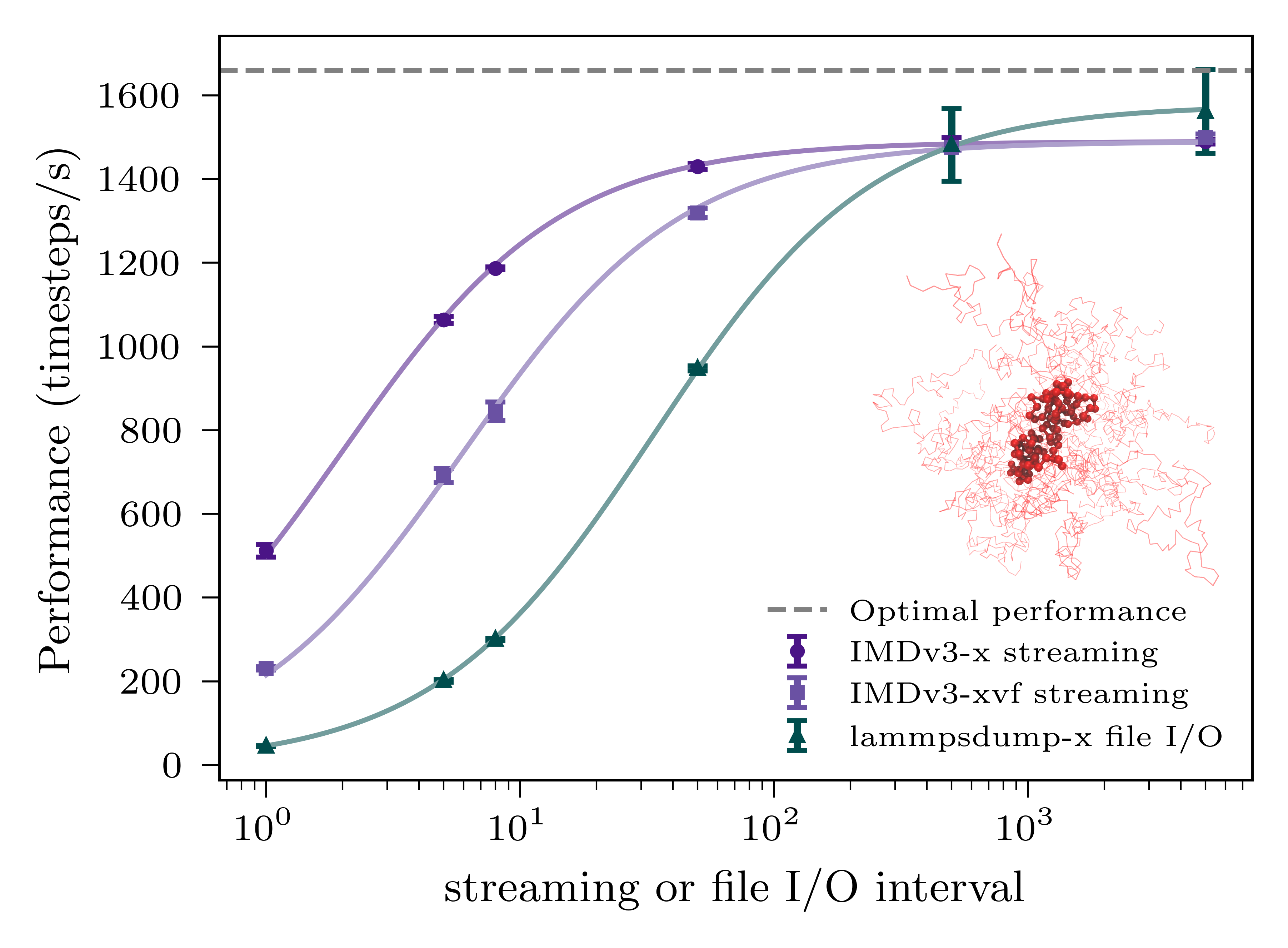}
  \caption{\textbf{Streaming is up to $\sim 10\times$ faster than file I/O in \software{LAMMPS}:} Comparing various streaming and file I/O scenarios in \software{LAMMPS} provides insights into the \software{LAMMPS} streaming implementation. For file I/O in \software{LAMMPS}, we only consider writing positions to a \software{LAMMPS} dump file (lammpsdump-x) due to disk space constraints and high runtimes. While streaming in \software{LAMMPS} (IMDv3-x) does provide $>10\times$ speed as compared to \software{LAMMPS} dump-x at lowest output intervals, the performance remains well below the optimal baseline, even at the highest streaming intervals considered. File I/O comes closer to the optimal baseline, indicating the constant per-step overhead associated with file I/O ($t^{(\text{x}, \text{lammpsdump}, 0)}_\text{IO}$) is lower than that for streaming ($t^{(0)}_\text{s}$) (see \Cref{tabSI:fit-params-all-engines} of the Supporting Information for more details).
  }
  \label{fig:LAMMPS-benchmarking}
\end{figure}

\begin{figure}[h!]
  \centering
  \includegraphics[width=\linewidth]{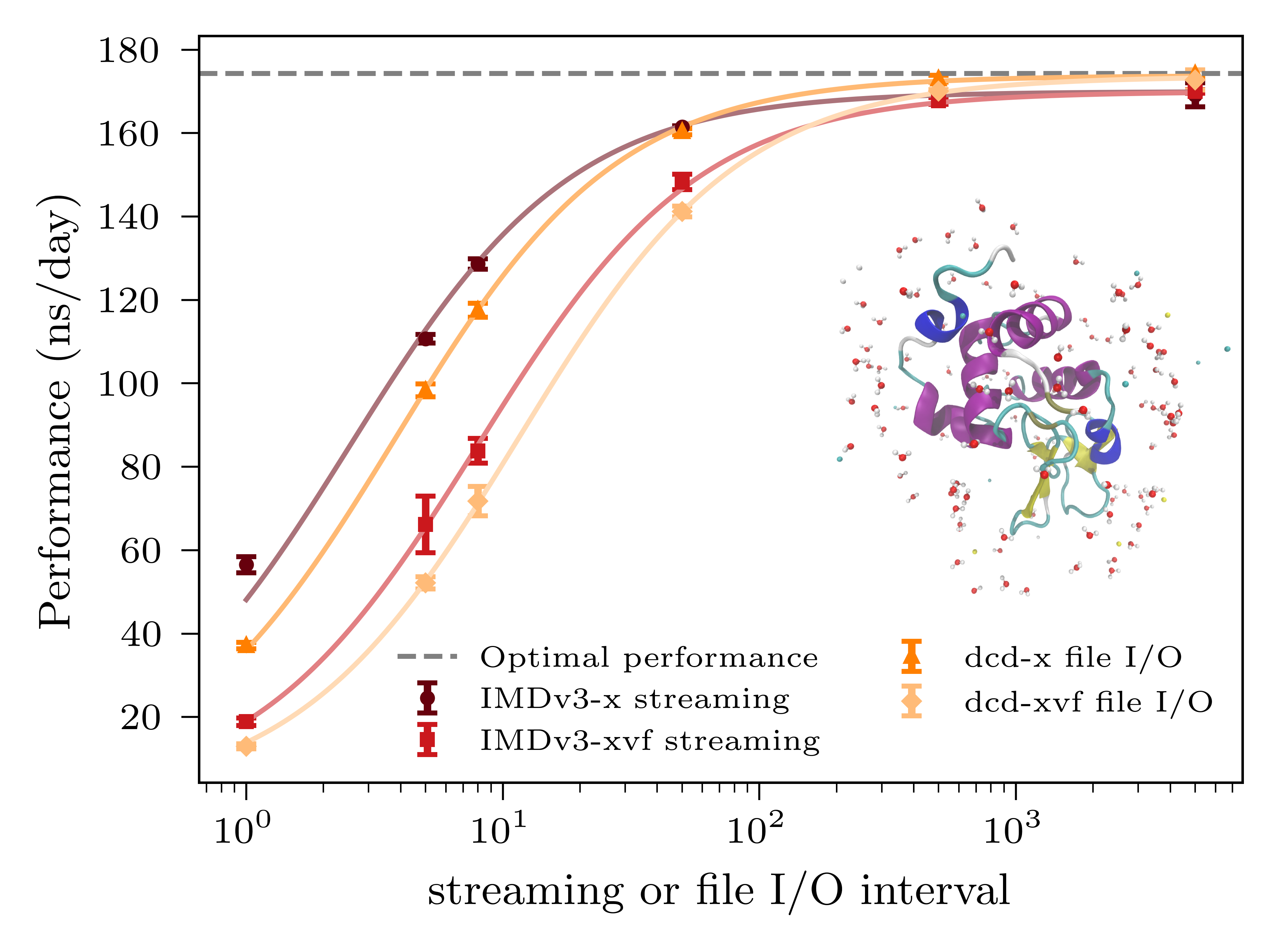}
  \caption{\textbf{Streaming is up to $\sim 1.5\times$ faster than file I/O in \software{NAMD}:} Performance with streaming and File I/O (to DCD files) was measured for positions only (IMDv3-x or dcd-x) and positions/velocities/forces (streaming-xvf or dcd-xvf). In \software{NAMD}, streaming provides a smaller advantage over file I/O at $\sim 1.5$ for the shortest streaming/file I/O intervals. This speedup reduces at higher intervals with both scenarios asymptotically trending to the baseline. Similar to \software{LAMMPS}, the streaming curves show some deviation here, indicating a larger per-step overhead, $t^{(0)}_\text{s}$ in the streaming implementation as compared to file I/O in \software{NAMD} (see \Cref{tabSI:fit-params-all-engines} of the Supporting Information for more details).
  }
  \label{fig:NAMD-benchmarking}
\end{figure}

The results are consistent with the general trend observed previously with \software{GROMACS}.
However, for both \software{LAMMPS} (\Cref{fig:LAMMPS-benchmarking}) and \software{NAMD} (\Cref{fig:NAMD-benchmarking}), we see a slowdown in performance when streaming frequently when compared to the optimal baseline.
Nevertheless, streaming continues to offer a significant speed advantage over file I/O, with a performance increase $>10 \times$ for short streaming intervals for \software{LAMMPS}.
As with \software{GROMACS}, we see that both data transfer schemes converge and plateau at higher data transmission/output frequencies (\Cref{fig:LAMMPS-benchmarking}).
For \software{NAMD}, streaming maintains a performance advantage with a speedup that is less dramatic but noticeable, with a $\sim 1.5$-fold increase in simulation performance when file I/O at every simulation time step is replaced by data streaming with \package{imdclient} (\Cref{fig:NAMD-benchmarking}).

For both \software{LAMMPS} and \software{NAMD}, the global model (\Cref{eq:master-sim-model}), describes the behaviors well for all streaming and file I/O scenarios considered.
Notably, for \software{LAMMPS}, the constant per-step cost of streaming information, $t^{(0)}_\text{s}$, is significant enough to cause a noticeable slowdown compared to the baseline even at higher streaming intervals.
This constant cost, $t^{(0)}_{\text{IO}}$ is still present but less significant for file I/O, producing the behavior observed in \Cref{fig:LAMMPS-benchmarking}.
For \software{NAMD}, the constant per-step costs for both streaming and file I/O are negligible, with streaming having a slightly higher cost, as seen in \Cref{fig:NAMD-benchmarking} and \Cref{tabSI:fit-params-all-engines} of the Supporting Information.

Unlike for \software{GROMACS}, both \software{LAMMPS} and \software{NAMD} suffer a significant drop in performance when streaming only positions at short intervals.
\software{GROMACS} shows almost no slowdown even when streaming at the shortest intervals.
This contrast can be attributed to the foundational IMD implementations (IMDv2) in each simulation engine that IMDv3 inherits.
The IMDv2 implementation in \software{GROMACS} \cite{IMDGMXGrubmuller} is highly optimized and compatible with GPU-accelerated MD simulations.
This was initially not the case for the IMDv2 implementations in \software{LAMMPS} and \software{NAMD}, which we modified to improve compatibility with GPU-accelerated simulation options (`kokkos' in \software{LAMMPS} and `GPU-resident-mode' in \software{NAMD}).

%However, the IMDv2 implementations in \software{LAMMPS} and \software{NAMD} needed significant modifications on our end to be made compatible with and support GPU-accelerated simulation options (like `kokkos` in \software{LAMMPS} and `GPU-resident-mode` in \software{NAMD}, respectively).

\section{Discussion}

The above IMDv3 streaming setup with the IMD-enabled streaming engine and \package{imdclient}, can be used to tackle many operational and logistical problems in molecular dynamics based research. In particular, streaming can be used for applications such as live monitoring of simulations, in-situ or on-the-fly analysis, or adaptive data sampling of simulation data.

In many scientific contexts, it is useful and prudent to monitor the progress of an experiment.
The same holds true for computational studies, where monitoring the progress of a simulation provides the potential for feedback control. 
In one of the most straightforward use cases, live monitoring can help terminate simulations that evolve incorrectly, saving time and computational resources. 
With IMDv3 streaming enabled, one can connect to any running simulation with an open port and use a client such as \package{imdclient} to receive live data on simulation progress. 
One can disconnect from the simulation at any time and control the resulting behavior of the simulation engine ({\em e.g.}, keep running in the background) with a specific IMDv3 message type (\codeinline{wait}).
The user can then reconnect again later to check on the progress of the simulation. This allows for a flexible and efficient way to monitor simulations.
More importantly, data streaming can be used for on-the-fly data analysis -- to calculate properties such as RMSD, RMSF, radial distribution functions, etc. -- and visualize them in real-time. Furthermore, given that the user has access to simulation data on-the-fly, it is straightforward to implement adaptive sampling strategies to process and analyze data at customizable time intervals or for a particular subset of the system. In practice, this could look like sampling the positions of a particular protein residue at logarithmic time intervals to capture its dynamic behavior on timescales spanning several orders of magnitude. In effect, our IMDv3 streaming setup provides a flexible and efficient framework to tackle many challenges in molecular simulations, which opens up new strategies for simulation data analysis.

\subsection{IMDv3 streaming in use}
\label{subsec:imdv3-apps}

In the following, we present practical usage examples of molecular dynamics simulations with IMDv3 streaming and highlight the benefits of streaming to overcome challenges posed by conventional approaches. 
We begin with the use of streaming to monitor running simulations -- doing so by intermittently connecting to a running simulation and calculating a physical property of interest to drive further decision making. 
In this example, we use \software{LAMMPS} and calculate the end-to-end distance of one of the polymers during a simulation of the system introduced in \Cref{subsec:imd-benchmarking} (see \Cref{fig:imd-applicationsa}). 
The use of IMD streaming here enables the user to monitor the simulation directly without having to write simulation data to a file and then perform analysis on it. 
This is particularly useful when performing equilibration runs (for large system sizes) and data production or storage is not needed. 
This can help the user decide whether the simulation is doing what it is supposed to do or whether it needs to be terminated. 
Further, as can be seen from the $t^{(0)}_{\text{s}}$ values in \Cref{tabSI:fit-params-all-engines} of the Supporting Information, there is a very low to negligible speed penalty associated with running a simulation in the background with just IMD turned on.
This makes IMDv3-streaming an efficient monitoring tool for various MD simulation scenarios.

Next, we show an analysis example that requires access to simulation data from previous simulation time steps.
To implement this, we store simulation data for a pre-determined number of simulation time steps in a ring buffer, which allows us to compute, {\em e.g.}, time correlation functions with a corresponding maximum delay time (here: $1$ ps).
As new data is received, the oldest simulation data are overwritten, thus only a fixed number of simulation data points are stored at any given time. In \Cref{fig:imd-applicationsb}, we simulate a system of $\sim 1400$ water molecules at $300$ K and $1$ bar in \software{GROMACS}, and calculate the translational velocity autocorrelation (VACF) over a fixed delay time window.
This VACF is recalculated and averaged continuously once the ring buffer has a full record over the past ps of simulation time. We perform the analysis for a single randomly selected water molecule in the system and observe how its time-averaged VACF converges with increasing simulation time to the overall ensemble average for the system (shown as reference in \Cref{fig:imd-applicationsb}).

Finally, we present a classic example of the usefulness of IMD streaming, where it facilitates the ability to track rare events in MD simulations. \Cref{fig:imd-applicationsd} shows a snapshot of a \software{NAMD} simulation that models ion transport through a membrane pore (more details can be found in \Cref{subsubsubsecSI:ion-current-membrane} of the Supporting Information). The ion current through this membrane is calculated by accessing live simulation data at high frequencies from an IMDv3 stream. As shown in \Cref{fig:imd-applicationsc}(inset), an analysis of the simulation trajectory at lower frequencies (longer time intervals) does not detect high frequency changes in the instantaneous current. More importantly, for longer time intervals, ambiguities arise between transport events through the membrane pore and diffusion events through the aqueous solvent across periodic boundaries of the simulation box. The use of IMDv3 streaming allows us to track such events at a high time resolution (\Cref{fig:imd-applicationsc}), allowing the user to unambiguously identify transport events.
We have provided a demo video describing the workflow to run the NAMD simulation and client script, calculating and visualizing ion currents from live MD data as part of the Supporting Information (see Video S1).

The scripts to run the simulation and client/analysis routines for the above applications, have been compiled in a GitHub repository (see \Cref{tabSI:software-availability} of the Supporting Information) and archived on Zenodo \cite{IMDv3appsThirumalaiswamy2026} for future reference. 

\begin{figure*}[th!]
\centering
\begin{captivy}{\includegraphics[width=\linewidth]{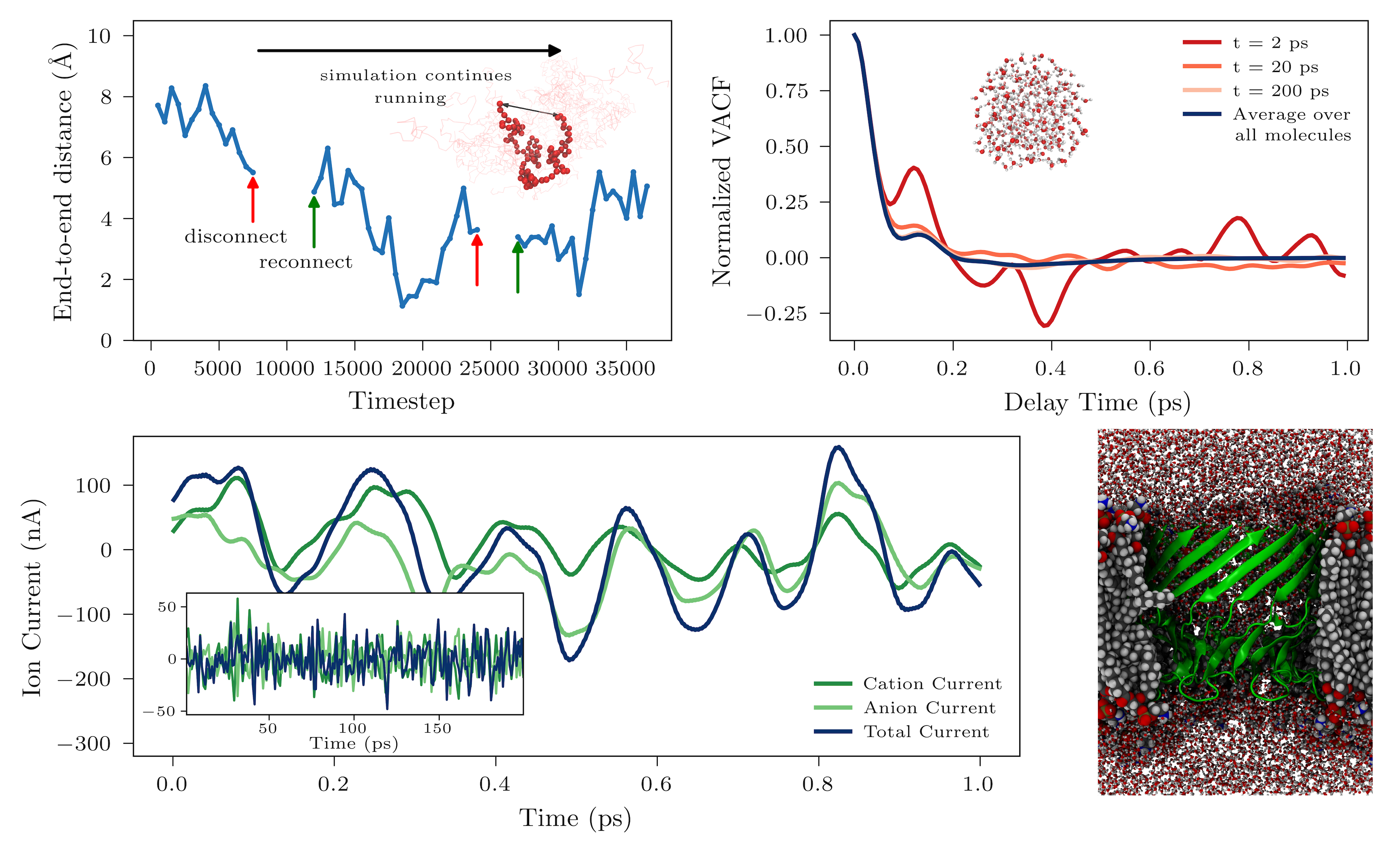}}
    \oversubcaption{0.07, 0.99}{}{fig:imd-applicationsa}
    \oversubcaption{0.57, 0.99}{}{fig:imd-applicationsb}
    \oversubcaption{0.07, 0.52}{}{fig:imd-applicationsc}
    \oversubcaption{0.76, 0.455}{}{fig:imd-applicationsd}
\end{captivy}
\caption{\textbf{Example applications of IMDv3 streaming: } 
(\textbf{a}) Intermittent monitoring of a live, running simulation using fully customizable quantities of interest, {\em e.g.}, here by calculating the end-to-end distance of a simulated polymer. The inset graphic shows the polymer simulation with one polymer selected for monitoring/analysis.
(\textbf{b}) Live monitoring of the convergence of a single-molecule velocity autocorrelation function (VACF) at different simulation times. The average VACF for the entire system at the end of the simulation ($t = 220 ps$).
(\textbf{c}) Monitoring of instantaneous ion currents through a membrane pore calculated by sampling ion positions at short time intervals (every time step) and longer time intervals (every $500$ timesteps) (inset).
(\textbf{d}) Graphical rendering of the membrane pore simulated in (\textbf{c}) (image produced with VMD \cite{HumphreyVMD1996}).
}
\label{fig:imd-applications}
\end{figure*}

\subsection{Streaming across the network}
\label{subsec:network}

We note here that the speedup advantage attributed to IMDv3-streaming can be affected by the quality of the network connection between the simulation engine and client. Network latency and bandwidth play a crucial role in information exchange across the IMD socket connection, indirectly affecting simulation engine speeds. In our benchmarking results, the use of a single node for both the simulation engine and client provides a high-speed interface without significant connection-based bottlenecks. This allows us to isolate and study the speedup enabled by IMDv3. In practice, we did not observe any slowdown when using a multi-node setup (simulation and client running on separate nodes) on a well-connected (Infiniband connection) supercomputing cluster like Arizona State University's SOL supercomputer \cite{SOLJennewein2023}. However, connecting to such a running simulation on a remote cluster with a local client (running on a personal computer) is negatively impacted by network connection quality, with network bandwidth and latency playing a limiting role in the ability to communicate via the TCP/IP socket connection.

\section{Conclusion}
\label{sec:conclusion}
In this study, we implemented a modified version of Interactive Molecular Dynamics {\em i.e.} IMDv3, that provides access to live simulation data and enables monitoring and \textit{in-situ} analysis of simulation properties. 
In particular, our implementation expands the information that a simulation engine can transfer as a stream and allows the user to configure which information should be included. 
Specifically, the user can specify any combination of atomic positions, velocities, and forces, periodic box dimensions, and energies. This information can be transmitted via the stream at a user-defined interval measured in simulation time steps ({\em i.e.}, transmission rate).
We have made this implementation available in \software{GROMACS}, \software{LAMMPS} and \software{NAMD}. To receive data streams in the IMDv3 format from these modified simulation engines, we have implemented the \package{imdclient} Python package, which presents the data time-frame by time-frame as Python data structures. 
Further, we have integrated a new reader class in \package{MDAnalysis} that reads and loads data provided by \package{imdclient} into \package{MDAnalysis} data structures, in a manner analogous to data input from a trajectory file. This enables the direct use of many built-in analysis methods available in \package{MDAnalysis}, to process the streamed simulation data.
Likewise, the new \package{MDAnalysis} reader class provides easy access to the streamed simulation data for any custom-built analysis.

We built our IMDv3 implementation directly on IMDv2, which has already been available in many simulation engines, and included code to send positions and energies to a client ({\em i.e.}, VMD) alongside the option to receive forces from the client. 
Our modified version of the code allows the user to switch between IMD versions, thus providing full backward compatibility with IMDv2. 
In addition to customizing the information that is being sent from the simulation engine to the client time-frame by time-frame (see above), our new protocol includes additional switches, {\em e.g.}, whether molecular coordinates are unwrapped (molecules are whole, {\em i.e.}, no covalent bonds are longer than half the box size but atoms can be located outside the periodic box) or wrapped (all atoms are within the boundaries of the periodic box, but covalent bonds may include 'jumps' across periodic boundaries).
On the receiver's end, our client implementation (\package{imdclient}), processes the data stream efficiently by housing the data temporarily in an internal buffer and using \codeinline{pause} and \codeinline{resume} commands to avoid any data loss.
Further, the modified protocol gives our client, \package{imdclient}, the ability to toggle the simulation's behavior to continue running or wait for a new connection when an existing IMD connection is severed.

The above-described characteristics and features of the IMDv3 protocol and implementation provide a flexible and robust framework for live-streaming of simulation data that has many potential use cases. 
The most obvious and generic use case would be to monitor live simulations by accessing critical simulation data and computing relevant order parameters. 
This could inform the user on the status of a specific simulation -- allowing decisions on whether to continue or terminate a specific simulation. 
The setting of \codeinline{wait} option in IMDv3 allows the user to do this intermittently by allowing the simulation to continue to run after a client connection is severed. 
In this case, the user can access live-simulation data at convenient intervals, essentially monitoring the simulation akin to an experiment. 
Furthermore, molecular dynamics simulations produce data on timescales that span many orders of magnitude. 
Information on fast fluctuations is often lost in long simulations because writing trajectory files at short time intervals both decreases simulation performance and drastically increases mass storage requirements.
With IMDv3, fast fluctuations can be analyzed \textit{in-situ} without needing to write high time resolution trajectory data to output files for post-hoc processing. 
Instead, trajectory output can be limited to low time resolution information, which reports on slow processes. 
Our integration of \package{imdclient} within \package{MDAnalysis} provides easy access to numerous built-in analysis tools that are easily extendable with user-defined code. 
In addition, it is straightforward to implement custom sampling strategies across various length scales and time scales, without having to run multiple copies of the simulation. 
Using the trajectory writer functions in \package{MDAnalysis}, the user can easily customize the trajectory output from a molecular dynamics simulation and include processing steps such as selecting a subsystem ({\em e.g.}, a protein without its solvent), applying rotational alignment transformations, {\em etc.}. 
It is also straightforward to sample system properties on logarithmic timescales or to generate custom output files for relevant system properties for record keeping using the ability of \package{MDAnalysis} to write most of the commonly used trajectory formats.

Overall, our IMDv3 implementation provides a framework that can transform data handling practices in molecular dynamics simulations. 
The ability to easily monitor simulations with user-defined parameters, to analyze fast fluctuations and dynamics without requiring large trajectory files, and to customize simulation output, greatly increases the flexibility of molecular simulation protocols. 
By building on the pre-existing IMDv2 framework, we were able to implement the IMDv3 protocol in multiple simulation engines and expect similar implementations to be straightforward elsewhere. 
A limitation of IMDv3 is the lack of support for communicating information back to the simulation engine, which would allow the user to modify simulation parameters \textit{in-situ} based on the intermediate analysis of live-simulation data. (Note: when IMDv2 is selected as the communication protocol during setup, external force communication to the MD engine remains fully supported). 
We hope to address some of these challenges in future versions of our protocol, which would further enhance the capabilities of interactive molecular dynamics. 
The inherent flexibility of IMD implementations in various simulation engines, alongside the modular nature of our client package (\package{imdclient}), makes such developments feasible.

\section*{Data and Software Availability}
Source codes and data discussed in this manuscript are freely available, as discussed in \Cref{subsec:software-avail}. 
\Cref{tabSI:software-availability} in the Supplementary Information contains links and DOI's for repositories of the used versions of simulation engine codes (before and after implementation of IMDv3), \package{imdclient} and \package{MDAnalysis}. Likewise, \Cref{tabSI:software-availability} in the Supplementary Information contains links and DOI's for repositories containing input files, scripts and output data for the benchmarks presented in \Crefrange{fig:GROMACS-benchmarking}{fig:NAMD-benchmarking} as well as the applications of IMDv3-streaming presented in \Cref{fig:imd-applications}.

\begin{acknowledgements}
This work is supported by the National Science Foundation (grant number OAC-2311372).
The authors acknowledge Research Computing at Arizona State University for providing high performance computing resources that have contributed to the research results reported within this work.
The authors thank Haochuan Chen (UIUC) and Axel Kohlmeyer (Temple University) for their help in implementing IMDv3 in NAMD and LAMMPS, respectively. 
We further thank Christopher Maffeo and Aleksei Aksimentiev (UIUC) for the initial idea to use streaming to monitor ion currents as well as input files and analysis routines for the nanopore simulation.
\end{acknowledgements}

\clearpage

%\bibliography{citations}

%%%%%%%%%% Merge with supplemental materials %%%%%%%%%%
\pagebreak
\widetext
\newpage
\begin{center}
\textbf{\large Supporting Information: Streaming Molecular Dynamics Simulation Data for On-the-fly Processing and Analysis}
\end{center}
%%%%%%%%%% Merge with supplemental materials %%%%%%%%%%
%%%%%%%%%% Prefix a "S" to all equations, figures, tables and reset the counter %%%%%%%%%%
\setcounter{equation}{0}
\setcounter{figure}{0}
\setcounter{table}{0}
\setcounter{section}{0}
\setcounter{page}{1}
\makeatletter
\renewcommand{\theequation}{S\arabic{equation}}
\renewcommand{\thefigure}{S\arabic{figure}}
\renewcommand{\thetable}{S\arabic{table}}
\renewcommand{\thesection}{S\arabic{section}}
\renewcommand{\thepage}{S\arabic{page}}
%%%%%%%%%% Prefix a "S" to all equations, figures, tables and reset the counter %%%%%%%%%%

\section{Terms}
\label{secSI:terms}

Below is a list of terms and their definitions as used in the main and SI text.

\begin{description}
   \item[IMD] Interactive Molecular Dynamics -- a mechanism to interact with a running molecular dynamics simulation through live data transfer via a TCP/IP connection with a user-side client application.
   \item[TCP/IP] Transmission Control Protocol/Internet Protocol - a suite of rules and protocols used to communicate between computers connected over a network architecture.
   \item[Socket] A software endpoint that connects a computer application to the underlying network hardware through the computer's network interface and allows for receiving and sending data over that network connection.
   \item[Port number] The socket uses a numerical identifier called a port number to identify a specific process or application running on a computer that is connected to the network. The complete socket address is defined by the IP address of the computer's network interface and the port number.
   \item[IMD protocol] The original application-level protocol developed by \citet{IMDoriginalStone2001}, that defines the type and format of data exchanged between the simulation engine and client.
   \item[IMDv1, IMDv2] Implementations in simulation engines based on the original IMD protocol
   \item[IMDv3 protocol] The enhanced protocol described in \Cref{subsec:imdv3-protocol}, that enables sending additional simulation information.
   \item[IMDv3] Implementation of the IMDv3 protocol in the simulation engines. We also use it to refer to the client implementation (\package{imdclient}) of the IMDv3 protocol.
   \item[Simulation engine] The producer of simulation data which listens for a receiver connection and sends it simulation data.
   \item[Producer] Producer of data - typically refers to the simulation engine in the context of IMD.
   \item[Client] The receiver of simulation data, connects to listening simulation engine and receives simulation data through a socket.
   \item[Receiver] Receiver of data - typically refers to the client in the context of IMD.
   \item[IMD frame] All information at a particular simulation time step that is streamed as IMD data constitutes an IMD frame.
   \item[IMD transmission rate] The step interval i.e. number of simulation steps in between each IMD frame that is streamed. For example, if the transmission rate is 1, every integration step in the simulation will be streamed. The term data transfer interval is used synonymously with transmission rate in this paper. 
   \item[IMD system] The subset of all atoms in the simulation for which the simulation engine can output IMD data and for which the receiver can send forces to the simulation engine to apply to.
   \item[IMD energy block] A data structure specific to IMD which contains the simulation step counter and information on the energy of the simulated system (not just the IMD system).
\end{description}

\section{IMDv3 streaming: Protocol, Implementations, and Usage}
\label{secSI:imd-streaming-protocol-impl-usage}

\subsection{IMDv3 protocol specification}
\label{subsecSI:imd-specification}

Here we specify the full IMDv3 protocol; this specification is also available as part of the imdclient documentation under \href{https://imdclient.readthedocs.io/en/latest/protocol_v3.html}{IMDv3 specs} \cite{IMDv3Woods2025}. 
Part of the Interactive Molecular Dynamics (IMD) approach is a protocol that dictates rules for communicating molecular simulation data through a socket, termed the \emph{IMD protocol}. It allows for two-way communication: via IMD, a simulation engine sends data to a receiver and the receiver can send forces and certain requests back to the simulation engine. 
This application-level protocol used in IMD defines the particular type and format of data being communicated.
We have developed a new version of the IMD protocol, termed IMDv3, capable of communicating additional simulation information relevant to most MD simulation-based scientific research.
Version numbers of IMD implementations are monotonically increasing integers. Previously available implementations, IMDv1 and IMDv2, were built on the original IMD protocol. The IMDv3 protocol described in this document inherits its name from its implementation's version number convention {\emph viz.} IMD version $3$ or IMDv3. The IMDv3 implementation in all $3$ simulation engines considered here builds upon existing IMDv2 code. It must be noted that the IMDv3 protocol includesupports bidirectional communication with force inputs from the receiver similar to its predecessor, the original IMDv2 protocol. However, currently the IMDv3 implementation in \software{LAMMPS} and \software{NAMD} does not support receiving forces from the client.

\begin{table}[h]
\centering
\begin{tabular}{|c|l|}
\hline
\textbf{IMD version} & \textbf{Protocol specification} \\
\hline
1 & Original IMD protocol - \textit{A system for interactive molecular dynamics simulation} \cite{IMDoriginalStone2001} \\
2 & Same original IMD protocol, with API implementation available \cite{IMDv2Grayson2013} \\
3 & New IMDv3 protocol, with implementations in \software{GROMACS}, \software{LAMMPS}, \software{NAMD} and client package, \package{imdclient} \\
\hline
\end{tabular}
\caption{IMD protocol versions and specifications}
\label{tabSI:imd-specs}
\end{table}

\subsection{IMD data transfer}
\label{subsecSI:imd-data-transfer}

Data transfer in IMD happens in the form of message packets whose composition is set in our implementation by the IMDv3 protocol. Below, we describe the IMDv3 protocol-defined packet types relevant to our implementations for the various simulation engines and the \package{imdclient} package.

\subsubsection{Packet Types}
\label{subsubsecSI:imd-packet-types}

In IMD-specific data communication, data is sent in the form of message packets, with each message typically containing two sub-packets viz. a header packet and a body packet.

A header packet is composed of $8$ bytes. The first $4$ bytes contain the header type and the next 4 bytes serve as a flexible slot for holding other information. The header type specifies the type of information the overall message packet carries. All bytes in header packets are provided in network (big) endian order except for the second $4$ bytes of the handshake packet. This is described in greater detail in the \codeinline{handshake} section.

\begin{verbatim}
Header:
   <val> (int32) Header type
   <val> (dtype) Data slot description
\end{verbatim}

Some header packets sent by the simulation engine have associated body packets that contain additional data, like atomic coordinates. These body packets vary in length and composition, but their bytes are always encoded in the native endianness of the machine running the simulation engine.

The list of header types for the header packet, as allowed by the IMDv3 protocol, are listed below

\begin{table}[h]
\centering
\begin{tabular}{|l|c|c|}
\hline
\textbf{Header type} & \textbf{32-bit integer enum value} & \textbf{Present in original IMD protocol} \\
\hline
\codeinline{disconnect}        & 0  & \checkmark \\
\codeinline{energies}          & 1  & \checkmark \\
\codeinline{coordinates}       & 2  & \checkmark \\
\codeinline{go}                & 3  & \checkmark \\
\codeinline{handshake}         & 4  & \checkmark \\
\codeinline{kill}              & 5  & \checkmark \\
\codeinline{md-communication}  & 6  & \checkmark \\
\codeinline{pause}             & 7  & \checkmark \\
\codeinline{transmission-rate} & 8  & \checkmark \\
\codeinline{io-error}          & 9  & \checkmark \\
\codeinline{session-info}      & 10 & $\times$ \\
\codeinline{resume}            & 11 & $\times$ \\
\codeinline{time}              & 12 & $\times$ \\
\codeinline{box}               & 13 & $\times$ \\
\codeinline{velocities}        & 14 & $\times$ \\
\codeinline{forces}            & 15 & $\times$ \\
\codeinline{wait}              & 16 & $\times$ \\
\hline
\end{tabular}
\caption{Header types in IMDv3 and their compatibility with the original IMD protocol}
\label{tabSI:imd-header-types}
\end{table}

Below, each header type and its associated body packet (if present) is described in detail.

\paragraph{Disconnect}

Sent from the receiver to the simulation engine any time after the \codeinline{session info packet} has been sent to indicate that the simulation engine should close the connected socket. Whether the simulation engine pauses execution until another connection is made is an implementation decision.

\begin{verbatim}
Header:
   0 (int32) Disconnect
   <val> (no type) Unused slot, any value acceptable
\end{verbatim}

\paragraph{Energies}

Sent from the simulation engine to the receiver each IMD frame if energies were previously specified for this session in the \codeinline{session info packet}.

\textbf{Note:} While the integration step is included in this packet, this is a result of inheriting the IMD energy block from the original IMD protocol. It is recommended to make use of the $64$-bit integer integration step value from the \codeinline{time packet} in analysis code instead.

\begin{verbatim}
Header:
   1 (int32) Energies
   1 (int32) Number of IMD energy blocks being sent

Body:
   <val> (int32) Current integration step of the simulation
   <val> (float32) Absolute temperature
   <val> (float32) Total energy
   <val> (float32) Potential energy
   <val> (float32) Van der Waals energy
   <val> (float32) Coulomb interaction energy
   <val> (float32) Bonds energy
   <val> (float32) Angles energy
   <val> (float32) Dihedrals energy
   <val> (float32) Improper dihedrals energy
\end{verbatim}

\paragraph{Coordinates}

Sent from the simulation engine to the receiver each IMD frame if coordinates were previously specified for this session in the \codeinline{session info packet}.

\begin{verbatim}
Header:
   2 (int32) Coordinates
   <n_atoms> (int32) Number of atoms in the IMD system

Body:
   <array> (float32[n_atoms * 3]) X, Y, and Z coordinates of each atom in the
                                  IMD system encoded in the order
                                  [X1, Y1, Z1, ..., Xn, Yn, Zn]
\end{verbatim}

\paragraph{Go}

Sent from the receiver to the simulation engine after the receiver receives the \codeinline{handshake} and \codeinline{session-info} packets. 

If the simulation engine does not receive this packet within 1 second of sending the handshake and session info packets, it should assume the receiver is incompatible. Whether the simulation engine exits or accepts another connection after this is an implementation decision.

\begin{verbatim}
Header:
   3 (int32) Go
   <val> (no type) Unused slot, any value acceptable
\end{verbatim}

\paragraph{Handshake}

Sent from the simulation engine to the receiver after a socket connection is established. Unlike other header packets, the last four bytes of this packet are provided in the native endianness of the sending simulation engine's hardware.

The receiver can use this packet to determine both the IMD version of the session and the endianness of the simulation engine. By providing the endianness of the machine running the simulation engine, the bulk of the data being sent in the session, i.e. the body packets, do not have to be swapped by the simulation engine before being sent, speeding up execution.

\begin{verbatim}
Header:
   4 (int32) Handshake
   3 (int32, unswapped byte order) IMD version used in session
\end{verbatim}

\paragraph{Kill}

Sent from the receiver to the simulation engine any time after the \codeinline{session-info} has been sent to request that the simulation engine stops execution of the simulation and exits. Whether or not the simulation engine honors this request is an implementation decision.

\begin{verbatim}
Header:
   5 (int32) Kill
   <val> (no type) Unused slot, any value acceptable
\end{verbatim}

\paragraph{MD Communication}

Sent from the receiver to the simulation engine any time after the \codeinline{session-info} has been sent to request that the forces in the body packet are applied to the atoms specified in the body packet. Whether or not the simulation engine honors this request is an implementation decision.

\begin{verbatim}
Header:
   6 (int32) MD Communication
   <n_atoms> (int32) Number of atoms in the IMD system to apply forces to

Body:
   <array> (int32[n_atoms]) Indices of atoms in the IMD system to apply forces to
   <array> (float32[n_atoms * 3]) The X, Y, and Z components of forces to be applied to
                                  the atoms at the indices specified in the above array
\end{verbatim}

The array of IMD system indices does not need to be monotonically increasing, meaning the indices can be "out of order". However, the index array cannot contain any index twice. Force vectors acting on the same index should be combined before being sent to the simulation engine to be applied.

Though this packet is sent by the receiver, the rule that all body packets are sent in the native endianness of the machine running the simulation engine still applies here. The receiver must use the endianness it gets from the \textit{handshake} and swap the endianness of the indices and forces if necessary before sending.

\paragraph{Pause}

Sent from the receiver to the simulation engine any time after the \codeinline{session-info} has been sent to request that the simulation engine pauses execution of the simulation until a \codeinline{resume} is sent. Pause is idempotent, meaning subsequent pause packets sent after the first one will have no effect.

\begin{verbatim}
Header:
   7 (int32) Pause
   <val> (no type) Unused slot, any value acceptable
\end{verbatim}

\textbf{Note:} In the original IMD protocol, pause acted as a toggle, meaning sending a pause packet twice would pause and then resume the simulation's execution. In IMDv3, the \codeinline{resume} packet is required to resume a paused simulation since pausing is idempotent.

\paragraph{Transmission rate}

Sent from the receiver to the simulation engine any time after the \codeinline{session-info} has been sent to change the IMD transmission rate. 

\begin{verbatim}
Header:
   8 (int32) Transmission rate
   <val> (int32) New transmission rate. Any value less than 1 will reset 
                 the transmission rate to its default value (configured
                 by the simulation engine)
\end{verbatim}

\paragraph{IO Error}

Never sent from one party to another during an IMD session. Can be used internally by the simulation engine or receiver to indicate an error has occurred.

\begin{verbatim}
Header:
   9 (int32) IO Error
   <val> (no type) Unused slot, any value acceptable
\end{verbatim}

\paragraph{Session info}

Sent from the simulation engine to the receiver after the handshake to provide metadata about the simulation.

\begin{verbatim}
Header:
   10 (int32) Session info
   7 (int32) Number of 1-byte configuration options in the body packet

Body:
   <val> (int8) Nonzero if time packets sent in each IMD frame
   <val> (int8) Nonzero if IMD energy block packets sent in each IMD frame
   <val> (int8) Nonzero if box packets sent in each IMD frame
   <val> (int8) Nonzero if coordinate packets sent in each IMD frame
   <val> (int8) Nonzero if coordinates wrapped into the simulation box.
                Meaningless if coordinates not sent in the session
   <val> (int8) Nonzero if velocity packets sent in each IMD frame
   <val> (int8) Nonzero if force packets sent in each IMD frame
\end{verbatim}

\paragraph{Resume}

Sent from the receiver to the simulation engine to resume execution after a pause.

\begin{verbatim}
Header:
   11 (int32) Resume
   <val> (no type) Unused slot, any value acceptable
\end{verbatim}

\paragraph{Time}

Sent from the simulation engine to the receiver each IMD frame if time packets were specified for the session.

\begin{verbatim}
Header:
   12 (int32) Time
   1 (int32) Number of time packets being sent

Body:
   <float64> dt for the simulation
   <float64> Current time of the simulation
   <int64> Current integration step of the simulation
\end{verbatim}

\paragraph{Box}

Sent from the simulation engine to the receiver each IMD frame if box packets were specified for the session.

\begin{verbatim}
Header:
   13 (int32) Box
   1 (int32) Number of simulation boxes being sent

Body:
   <float32[9]> Triclinic box vectors for the simulation encoded in order
                [ABC], where
                A = (aX, aY, aZ),
                B = (bX, bY, bZ),
                C = (cX, cY, cZ)
\end{verbatim}

\paragraph{Velocities}

Sent from the simulation engine to the receiver each IMD frame if velocities were specified for the session.

\begin{verbatim}
Header:
   14 (int32) Velocities
   <n_atoms> (int32) Number of atoms in the IMD system

Body:
   <float32[n_atoms * 3]> X, Y, and Z components of velocities of each atom,
                         encoded as [Vx1, Vy1, Vz1, ..., Vxn, Vyn, Vzn]
\end{verbatim}

\paragraph{Forces}

Sent from the simulation engine to the receiver each IMD frame if forces were specified for the session.

\begin{verbatim}
Header:
   15 (int32) Forces
   <n_atoms> (int32) Number of atoms in the IMD system

Body:
   <float32[n_atoms * 3]> X, Y, and Z components of forces on each atom,
                         encoded as [Fx1, Fy1, Fz1, ..., Fxn, Fyn, Fzn]
\end{verbatim}

%-------

\paragraph{Wait}

Sent from the receiver to the simulation engine any time after the \codeinline{session info packet} has been sent to request that the simulation engine modify its waiting behavior mid-simulation either from blocking to non-blocking or vice versa. Whether or not the simulation engine honors this request is an implementation decision.

Regardless of whether this packet is accepted, the simulation engine will have an initial waiting behavior which applies to the beginning of the simulation:

\begin{enumerate}
  \item Blocking: Wait until a receiver is connected to begin execution of the simulation
  \item Non-blocking: Begin the simulation regardless of whether a receiver is connected and continuously check on the listening socket for a receiver attempting to connect
\end{enumerate}

The simulation engine's waiting behavior also applies when a receiver disconnects mid-simulation:

\begin{enumerate}
  \item Blocking: Pause simulation execution and wait until a receiver is connected to resume execution
  \item Non-blocking: Continue execution, continuously checking on the listening socket for a receiver attempting to connect
\end{enumerate}

\begin{verbatim}
Header:
   16 (int32) Wait
   <val> (int32) Nonzero to set the simulation engine's waiting behavior to blocking, 0
                 to set the simulation engine's waiting behavior to non-blocking
\end{verbatim}

\textbf{Note:} The purpose of this packet is to allow a receiver to monitor the first *n* frames of a simulation and then disconnect without blocking the continued execution of the simulation.

\subsubsection{Packet order}
\label{subsubsecSI:packet-order}

After the simulation engine sends the \codeinline{handshake} and \codeinline{session info} packets to the receiver and gets back a \codeinline{go} signal, it begins sending simulation data via IMD. The data within each IMD frame is always sent in the same, fixed order:

\begin{enumerate}
  \item Time
  \item Energy block
  \item Box
  \item Coordinates
  \item Velocities
  \item Forces
\end{enumerate}

If the simulation engine is configured to send only a strict subset of all available data packets, the fixed order of the list still applies to the remaining packets in the session.

\textbf{Note:} In IMDv3 implementations, the simulation engine and client require that all packets specified in session info must be sent for every IMD frame and in the same order. In contrast, IMDv2 allowed any packet order and different packets per frame.

\subsection{Units}
\label{subsecSI:units}

The units in IMDv3 are fixed. The simulation engine must convert values into these units before sending them through the socket. The receiver must also convert forces it sends back to the simulation engine into these units.

\begin{center}
\begin{tabular}{|l|l|l|}
\hline
Measurement & Unit & Shorthand Notation \\
\hline
Length & Angstrom & Å \\
Velocity & Angstrom per picosecond & Å/ps \\
Force & Kilojoules per mole per angstrom & kJ/(mol·Å) \\
Time & Picosecond & ps \\
Energy & Kilojoules per mole & kJ/mol \\
\hline
\end{tabular}
\end{center}

\subsection{IMD port number}
\label{imdSI-port-number}

The preferred port for IMD communication is 8888, but the simulation engine may freely specify the port at which it listens for a receiver.

\section{IMDv3 implementations: Producer and Receiver}
\label{secSI:imdv3-implementations}

Our IMDv3-streaming setup requires implementations of the IMDv3 protocol and its functionality on the producer and receiver ends of the IMD connection. To achieve that, we provide IMDv3 functions and methods in three popular simulation engines: \software{GROMACS}, \software{NAMD}, and \software{LAMMPS} that are accessible via settings in their input files. On the receiver end, we provide a Python package, \package{imdclient}, that is capable of receiving and processing IMDv3 data streams from any producer following the IMDv3 protocol. Finally, we implement a reader within \package{MDAnalysis}, that allows users to load IMDv3 data streams into \package{MDAnalysis} data structures for convenient access and analysis.

\subsection{Simulation engine implementations}
\label{subsecSI:md-engine-implementation}

We have currently implemented IMDv3 streaming in three popular molecular dynamics simulation engines: \software{GROMACS} \cite{GROMACSPall2020}, \software{NAMD} \cite{NAMDPhillips2020}, and \software{LAMMPS} \cite{LAMMPSThompson2022}. The details of the availability of these implementations are discussed in \Cref{subsec:software-avail}, with links tabulated in \Cref{tabSI:software-availability}.

Below we provide detailed instructions on how to use and enable IMDv3 functionality in each of the three simulation engines.

\subsubsection{\software{GROMACS}}
\label{subsubsecSI:gromacs-implementation}
To run IMDv3 in \software{GROMACS} one has to make the following changes to the molecular dynamics parameter ( \codeinline{*.mdp}) input file:

\begin{lstlisting}[style=mdp]
; Required IMD settings
IMD-group        = System   
; Atom group to stream (System = all atoms)

IMD-nst          = 100      
; Send data every 100 steps
IMD-version      = 3        
; Use IMDv3 protocol (2/3) - defaults to 2
    
; Data types and specifications - what to stream
IMD-time         = Yes       
; Stream timing information (Yes/No)
IMD-energies     = Yes       
; Stream energy data (Yes/No)
IMD-box          = Yes       
; Stream box dimensions (Yes/No)
IMD-coords       = Yes       
; Stream coordinates (Yes/No)
IMD-vels         = Yes       
; Stream velocities (Yes/No)
IMD-forces       = Yes       
; Stream forces (Yes/No)
IMD-unwrap       = Yes       
; Coordinate processing before streaming - Unwrap coordinates across PBC (Yes/No)
\end{lstlisting}

\software{GROMACS} turns IMD functionality on/off using the \codeinline{IMD-group} input setting in the \codeinline{*.mdp} file. 
Further, \software{GROMACS} requires the user to define additional IMD-related variables, such as the port number and whether the simulation should wait for a client connection before starting, on the command line at runtime:

\begin{lstlisting}[style=bashstyle]
#!/bin/bash
gmx grompp -f input.mdp -c conf.gro -p topol.top -o run.tpr

gmx mdrun -v -deffnm run -imdport 8888 -imdwait
# gmx mdrun -gmx_flags -imdport <port number> -imdwait
# -imdport: port number for socket connection (default: 8888)
# -imdwait: optional flag to wait for client connection before starting
\end{lstlisting}

\subsubsection{LAMMPS}
\label{subsubsecSI:lammps-implementation}
In \software{LAMMPS}, one adds the following line to the input file to enable IMDv3 functionality:

\begin{lstlisting}[style=lammps]
# IMD configuration

# fix ID group-ID imd <port> version <2/3> nowait <on/off> trate <arg> time <yes/no> box <yes/no> coordinates <yes/no> velocities <yes/no> forces <yes/no> unwrap <yes/no>
fix imdv3 all imd 8888 nowait off trate 100 version 3 time yes box yes coordinates yes velocities yes forces yes unwrap yes

# Parameters explained:
# ID, group-ID : user-assigned fix name and ID of group of atoms fix applies to
# imd: fix style name to enable IMD
# 8888: port number for connection
# nowait off: Waits for client connection before starting simulation (on/off)
# trate 100: Send data every 100 timesteps (transmission rate)
# version 3: IMD protocol version (2/3), defaults to 2
# time yes: Send timing information (yes/no)
# box yes: Send box dimensions (yes/no)
# coordinates yes: Send atomic coordinates (yes/no)
# velocities yes: Send velocity data (yes/no)
# forces yes: Send force data (yes/no)
# unwrap yes: Unwrap coordinates across PBC (yes/no)
# Note: Energy streaming is not supported in \software{LAMMPS} IMD fix

\end{lstlisting}

Then, one can run \software{LAMMPS} as usual.

\subsubsection{\software{NAMD}}
\label{subsubsecSI:namd-implementation}
Finally, for \software{NAMD} one can similarly edit the input file with the following settings:

\begin{lstlisting}[style=namd]
# Required IMD settings
IMDon                   yes            ; Enable IMD functionality (yes/no)
IMDport                 8888           ; Port number for socket connection
IMDwait                 on             ; Wait for client connection before starting simulation (on/off)
IMDfreq                 100            ; Send data every 100 steps (transmission rate)
IMDversion              3              ; IMD protocol version (2/3)

# Data types - what to send
IMDsendTime             yes            ; Send timing information (yes/no)
IMDsendEnergies         yes            ; Send energy information (yes/no)
IMDsendBoxDimensions    yes            ; Send simulation box data (yes/no)
IMDsendPositions        yes            ; Send coordinates (yes/no)
IMDsendVelocities       yes            ; Send velocity data (yes/no)
IMDsendForces           yes            ; Send force data (yes/no)

# Coordinate processing
IMDwrapPositions        no             ; Whether to wrap coordinates into periodic simulation box (yes/no)

\end{lstlisting}

\software{NAMD} can be run as usual using the modified input file.

\textbf{Note}: It must be noted that currently the IMDv3 implementation in \software{LAMMPS} and \software{NAMD} does not support receiving and applying forces from the client.

\subsection{\package{imdclient} implementation}
\label{subsecSI:imdclient-implementation}

Our client implementation for the IMDv3 protocol is a Python package, \package{imdclient}, which is capable of receiving and processing IMDv3 streamed data as described in \Cref{subsubsec:imdclient-implementation}. Below we describe the implementation architecture in detail. At the time of writing, the latest release of the package is \href{https://github.com/Becksteinlab/imdclient/releases/tag/v0.2.3}{\codeinline{v0.2.3}}.

With the goal of reducing wasteful, throughput-reducing network round-trips between receiver and simulation engine where the receiver repeatedly advertises a window size of 0 in its socket receive buffer and the simulation engine repeatedly responds with exponentially-time-gapped zero-window probes inherent to TCP \cite{TCPEddy2022}, imdclient implements an architecture with an application-level buffer (\codeinline{IMDFrameBuffer}) and automatic pausing/resuming on the trajectory data stream based on high and low fill watermarks on the \codeinline{IMDFrameBuffer}.

To achieve this, imdclient is implemented as a two-threaded, single-producer/single-consumer architecture. The producer thread (\codeinline{IMDProducerV3}) reads IMDv3 packets continuously from the stream, formats and packages them into a frame-holding class (\codeinline{IMDFrame}), and then pushes these full \codeinline{IMDFrame}s into a queue of full frames ready for analysis managed by the \codeinline{IMDFrameBuffer} class. The consumer thread (\codeinline{IMDClient}) is the class with the user-facing API (\Cref{tabSI:api-methods-tabular}). Whenever the \codeinline{IMDClient}'s \codeinline{get\_imdframe()} method is called, a full \codeinline{IMDFrame} is popped from the \codeinline{IMDFrameBuffer}'s full queue for analysis and the previously analyzed frame is added to the \codeinline{IMDFrameBuffer}'s empty queue for re-filling by the \codeinline{IMDProducerV3}. This process is shown visually in \Cref{fig:imdclient-arch}.

If the producer thread fills a high proportion of the \codeinline{IMDFrameBuffer}'s available \codeinline{IMDFrame}s, it will send a pause signal to the simulation engine until it has 1. emptied the socket recv buffer of frames that have already arrived and 2. detected that the consumer thread has popped the number of full frames down below the low watermark. These watermark levels are configurable via keyword arguments '\codeinline{pause\_empty\_proportion}' and '\codeinline{unpause\_empty\_proportion}' described in \Cref{tabSI:api-methods-tabular}.

To avoid either thread waiting infinitely, the \codeinline{IMDFrameBuffer} allocates Python \codeinline{threading.Condition} locks acquirable by the producer and consumer threads which allow them to be notified when the other thread has stopped and handle shutdown gracefully. For example, if no frames are currently available in the \codeinline{IMDFrameBuffer}'s full queue when the consumer thread attempts to pop one, it will call \codeinline{wait()} on a condition lock and any exception or expected stop in the producer thread will wake up the consumer thread so that it doesn't wait forever on a stopped producer.

\setcellgapes{0pt}
\makegapedcells

\begin{table}[h!]
\centering
\caption{API methods and their behaviors}
\label{tabSI:api-methods-tabular}
\begin{tabular}{|p{0.35\linewidth}|p{0.35\linewidth}|p{0.30\linewidth}|}
\hline
\textbf{Method name} & \textbf{Returns} & \textbf{Raises} \\
\hline
\begin{tabular}[t]{@{}l@{}}
\codeinline{\_\_init\_\_(}\\
\codeinline{host: string,}\\
\codeinline{port: int,}\\
\codeinline{n\_atoms: int,}\\
\codeinline{socket\_bufsize: int} \\ 
\hspace{26 mm}\codeinline{(optional),}\\
\codeinline{timeout: int (optional),}\\
\codeinline{continue\_after\_disconnect: bool}\\
\hspace{45 mm}\codeinline{(optional),}\\
\codeinline{buffer\_size: int (optional),}\\
\codeinline{pause\_empty\_proportion: float}\\
\hspace{39 mm}\codeinline{(optional),}\\
\codeinline{unpause\_empty\_proportion: float} \\
\hspace{42 mm}\codeinline{(optional)}\\
\codeinline{)}
\end{tabular} & 
\codeinline{IMDClient} object with IMDv3 connection to '\codeinline{host:port}'. 

'\codeinline{n\_atoms}' is used in conjunction with the \codeinline{IMDSessionInfo} object to define the size of the individual \codeinline{IMDFrame}s used in the \codeinline{IMDFrameBuffer} for the data types present in the stream.

'\codeinline{buffer\_size}' (defaults to 10MB) defines the maximum amount of memory used by the \codeinline{IMDFrameBuffer} to allocate \codeinline{IMDFrame}s.

\codeinline{'socket\_bufsize}' (defaults to OS default) sets the \codeinline{IMDClient}'s socket recv buffer size. 

'\codeinline{timeout}' (defaults to 5) defines the number of seconds after which \codeinline{IMDProducerV3} will assume the simulation has ended after not receiving a TCP packet.

'\codeinline{continue\_after\_disconnect}' (default '\codeinline{None}') can change the simulation engine's waiting behavior from its initial configuration after the client disconnects. If '\codeinline{True}', the client will attempt to change the simulation engine's waiting behavior to non-blocking. If '\codeinline{False}', the client will attempt to change it to blocking. If '\codeinline{None}', the client will not attempt to change the simulation engine's behavior. 

'\codeinline{pause\_empty\_proportion}' sets the lower threshold proportion of the \codeinline{IMDFrameBuffer}'s total number of \codeinline{IMDFrame}s that must be empty before the simulation is paused (defaults to 0.25).

'\codeinline{unpause\_empty\_proportion}' sets the proportion of the \codeinline{IMDFrameBuffer}'s \codeinline{IMDFrame}s that must be empty before the simulation is unpaused (defaults to 0.5).  & 
\codeinline{ConnectionRefusedError} if a TCP connection cannot be established with the simulation engine

\codeinline{ConnectionError} if the handshake packet is not received

\codeinline{ValueError} if another packet is received before the handshake packet

\codeinline{ValueError} if an incompatible IMD version number is received in the handshake packet

\codeinline{ValueError} if an IMDSessionInfo packet is not received after the handshake packet

\codeinline{ValueError} if the IMDSessionInfo's length field is not 7

\codeinline{ValueError} if \newline \codeinline{pause\_empty\_proportion} or \newline \codeinline{unpause\_empty\_proportion} \newline is not in the range [0,1]

\codeinline{ValueError} if \codeinline{buffer\_size} is too small to hold a single IMDFrame given IMDSessionInfo and \codeinline{n\_atoms} \\
\hline
\codeinline{get\_imdframe()} & \codeinline{IMDFrame} object with fields '\codeinline{time}', '\codeinline{dt}', '\codeinline{step}', '\codeinline{energies}', '\codeinline{box}', '\codeinline{positions}', '\codeinline{velocities}', and '\codeinline{forces}'. Each field not present in the stream is set to \codeinline{None} & \codeinline{EOFError} if there are no more frames to read from the stream \\
\hline
\codeinline{get\_imdsessioninfo()} & \codeinline{IMDSessionInfo} object with all information obtained from the session info packet (booleans '\codeinline{time}', '\codeinline{box}', '\codeinline{positions}', '\codeinline{velocities}', and '\codeinline{forces}') as well as the protocol version (integer '\codeinline{version}') (3) and endianness of the simulation engine (string '\codeinline{endianness'},  encoded as the string ">" for big and "<" for little) & \\
\hline
\codeinline{stop()} & \codeinline{None}, stops the client and closes the connection & \\
\hline
\end{tabular}
\end{table}

\begin{table}[h!]
\centering
\caption{Protocol packet types and corresponding \package{imdclient} functions}
\label{tabSI:protocol-imdclient-mapping}
\renewcommand{\arraystretch}{1.2}
\begin{tabular}{|p{3.6cm}|p{8cm}|p{5cm}|}
\hline
\textbf{Protocol Packet Type} & \textbf{\package{imdclient} Function/Data structure} & \textbf{Notes} \\
\hline
\codeinline{handshake} & Internal handshake handling & Automatic during connection \\
\codeinline{session-info} & class \codeinline{IMDSessionInfo} & Accessible via function \newline \codeinline{IMDClient.get\_imdsessioninfo} \\
\codeinline{go} & Internal go signal & Sent automatically \\
\codeinline{disconnect} & function \codeinline{IMDClient.stop} & Closes connection \\
\codeinline{kill} & Not implemented & Attempts to stop connected simulation engine \\
\codeinline{pause} & Internal pause handling & Automatic buffer management \\
\codeinline{resume} & Internal resume handling & Automatic buffer management \\
\codeinline{wait} & function \codeinline{IMDClient.\_\_init\_\_} keyword arg \newline '\codeinline{continue\_after\_disconnect}' & See \Cref{tabSI:api-methods-tabular} for behavior \\
\codeinline{transmission-rate} & Not implemented & N/A \\
\hline
\codeinline{time} & class attributes \codeinline{IMDFrame.time}, \codeinline{IMDFrame.dt}, \newline \codeinline{IMDFrame.step} & Accessible via function \newline \codeinline{IMDClient.get\_imdframe} \\
\codeinline{energies} & \codeinline{IMDFrame.energies} & Accessible via \newline \codeinline{IMDClient.get\_imdframe} \\
\codeinline{box} & \codeinline{IMDFrame.box} & Accessible via \newline \codeinline{IMDClient.get\_imdframe} \\
\codeinline{coordinates} & \codeinline{IMDFrame.positions} & Accessible via \newline \codeinline{IMDClient.get\_imdframe} \\
\codeinline{velocities} & \codeinline{IMDFrame.velocities} & Accessible via \newline \codeinline{IMDClient.get\_imdframe} \\
\codeinline{forces} & \codeinline{IMDFrame.forces} & Accessible via \newline \codeinline{IMDClient.get\_imdframe} \\
\hline
\codeinline{md-communication} & Not implemented & Force feedback not implemented \\
\codeinline{io-error} & Not implemented & N/A \\
\hline
\end{tabular}
\end{table}

\section{Benchmarking IMDv3 streaming}
\label{secSI:benchmarking-imdv3-streaming}

Below, we provide more details on how we benchmarked the IMDv3 protocol implementation for the three simulation engines \software{GROMACS}, \software{LAMMPS}, and \software{NAMD}.

\subsection{GROMACS}
\label{subsecSI:benchmarking-gromacs}

As discussed in the main text, we run the IMDv3-modified and unmodified versions of \software{GROMACS}, varying the number of cores used, over a combination of thread-MPI tasks and OpenMP threads. We find that these two versions consistently match each other across these various settings, with both producing optimum performance when using $24$ cores with a single thread MPI task and $24$ OpenMP threads.
One may notice that under resources-constrained situations, one can always use say $12$ cores to achieve similar performance. For our benchmarking, we proceed to use the optimum configuration of $24$ cores.

\begin{figure}[h!]
  \centering
  \includegraphics[scale=1]{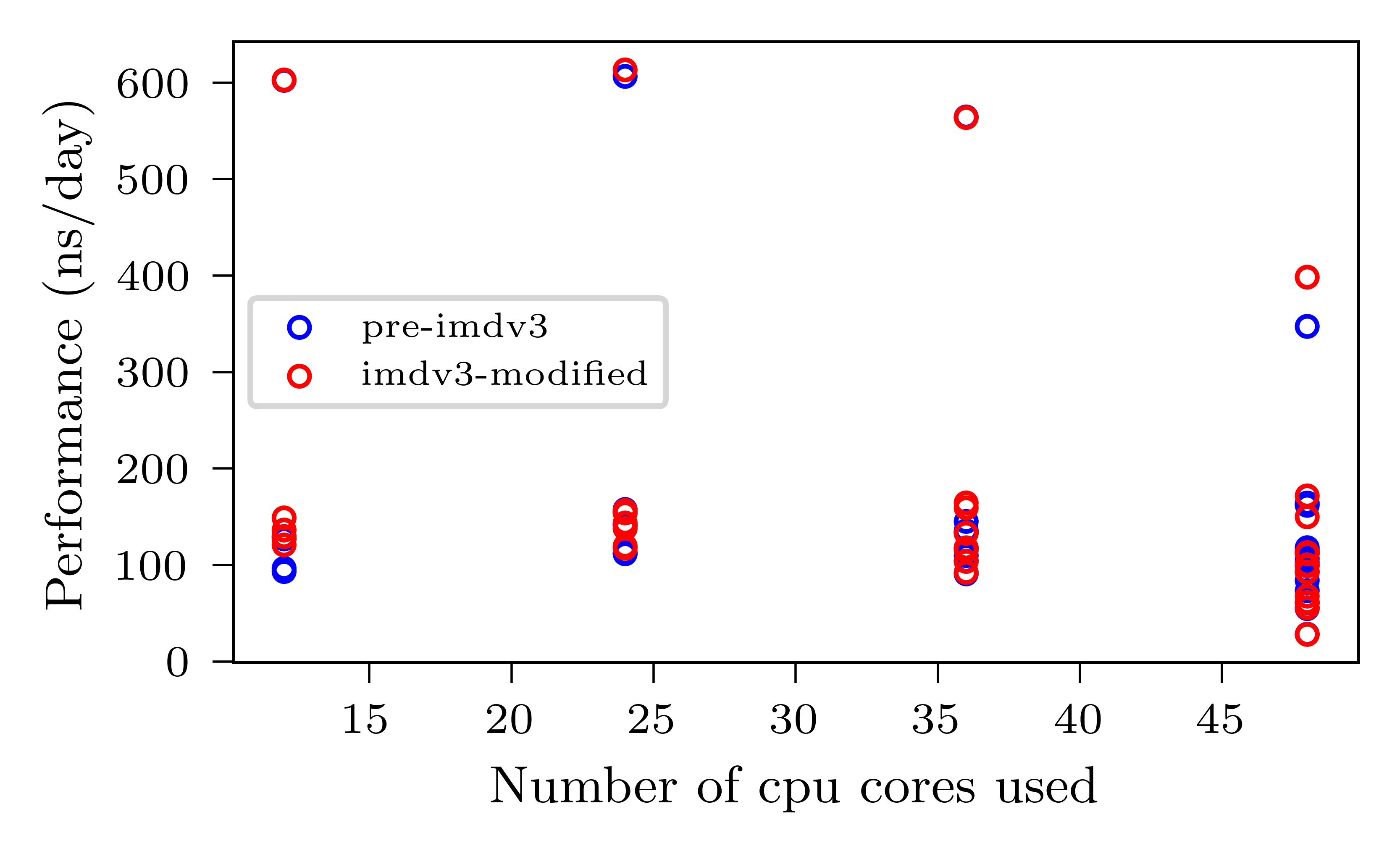}
   \caption{\textbf{Optimizing the performance of \software{GROMACS}:} We run $2$ versions of \software{GROMACS} viz. IMDv3-modified (`imdv3') and the pre-IMDv3 (vanilla) version over a fixed set of computational resources. We observe that at $24$ cores with $1$ thread-MPI task and $24$ OpenMP threads, \software{GROMACS} produces the best performance. This is consistent between the $2$ versions tested. The various data points at each core count represent different combinations of thread-MPI and OpenMP settings. However, using multiple thread-MPI tasks produces sub-optimal performance $\leq 200$ ns/day, compared to optimal speeds when using a single thread-MPI task.}
  \label{figSI:GROMACS-optimization}
\end{figure}

\subsection{LAMMPS}
We follow a similar optimization approach for \software{LAMMPS}, and test out various OpenMP, MPI and GPU based settings. The `kokkos'-mode in \software{LAMMPS} consistently gives best performance for the system of interest {\em i.e.}, polymers in a box. Running under the `kokkos' mode, we sweep across a range of cores (threads) used. Further, we consider both the IMDv3-modified (`imdv3') and pre-IMDv3 (`vanilla') versions of \software{LAMMPS} and find that optimal performance is observed when using $8$ cores (threads). Here, we note that unlike \software{GROMACS}, since IMDv3-based changes were implemented in multiple stages, the IMDv3-modified version contains multiple non-IMDv3 based optimizations from the `develop' branch of the official repository. This gives rise to consistent differences in speed between the $2$ versions, though not very significant.

\begin{figure}[h!]
  \centering
  \includegraphics[scale=1]{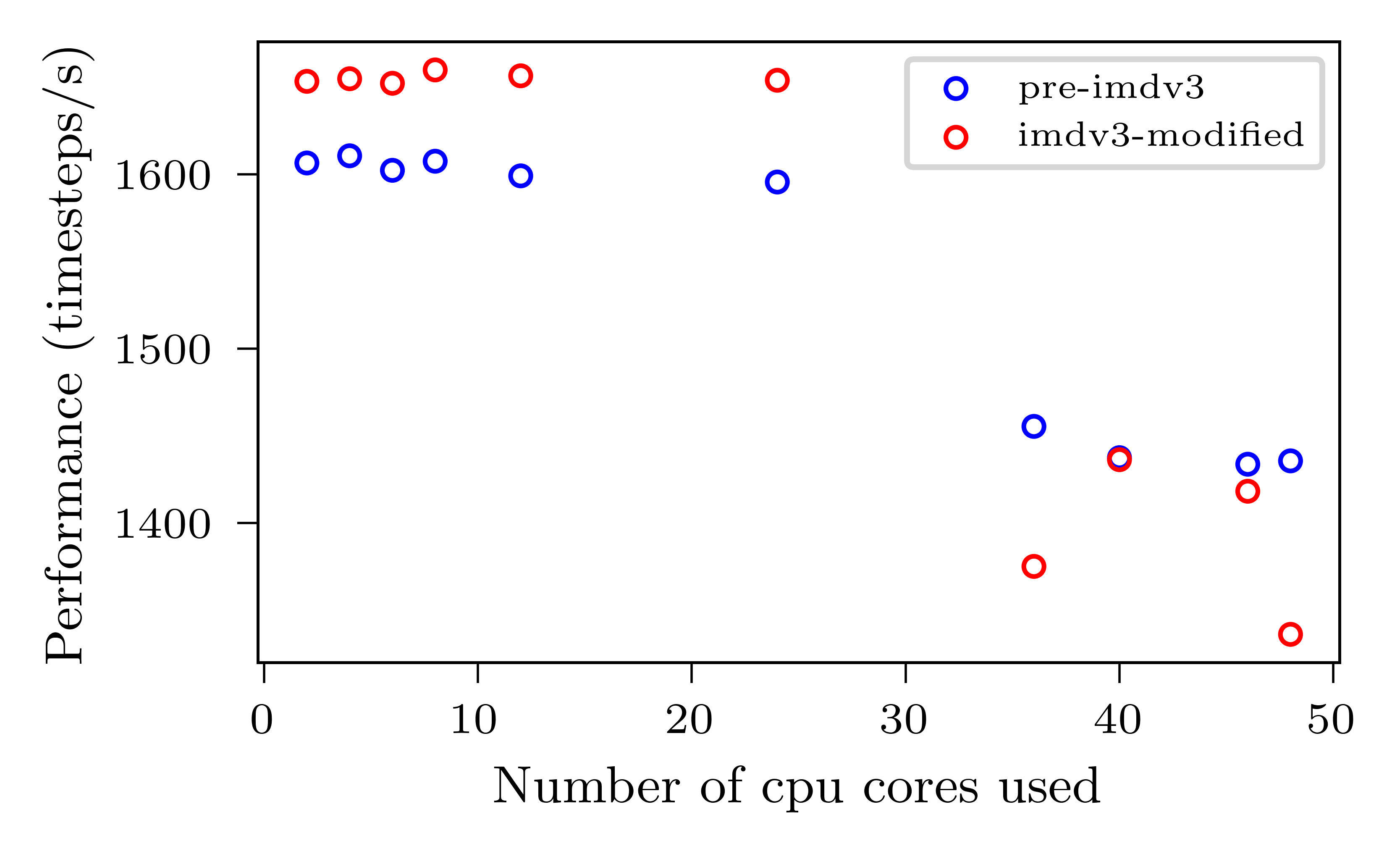}
  \caption{\textbf{Optimizing the performance of \software{LAMMPS}:} We run $2$ versions of \software{LAMMPS} viz. IMDv3-modified (`imdv3') and the pre-IMDv3 (`vanilla') version over a fixed set of computational resources. We observe that at $8$ cores (threads) under the `kokkos' mode gives best performance. This is consistent between the $2$ versions tested, despite noted differences in the $2$ versions as discussed in the main text and above.}
  \label{figSI:LAMMPS-optimization}
\end{figure}

\subsection{NAMD}

For \software{NAMD}, we again consider $2$ versions {\em viz.} IMDv3-modified (`imdv3') and pre-IMDv3 (`vanilla'), and find that the smp-MPI configuration with the `GPU resident mode' enabled gives the best performance. Like \software{LAMMPS}, the IMDv3-modified version of \software{NAMD} contains multiple non-IMDv3-based optimizations from the `main' branch of the official repository, which gives rise to consistent differences in speed between the $2$ versions. We sweep through different cores (worker threads) used, and find an optimum at $35$ threads. It must be noted here that under the smp-MPI mode, \software{NAMD} uses an additional communication thread by default, so total count of optimal cores would be $36$. However, for our final benchmarking we show results for $45$ ($+1$) threads instead.

\begin{figure}[h!]
  \centering
  \includegraphics[scale=1]{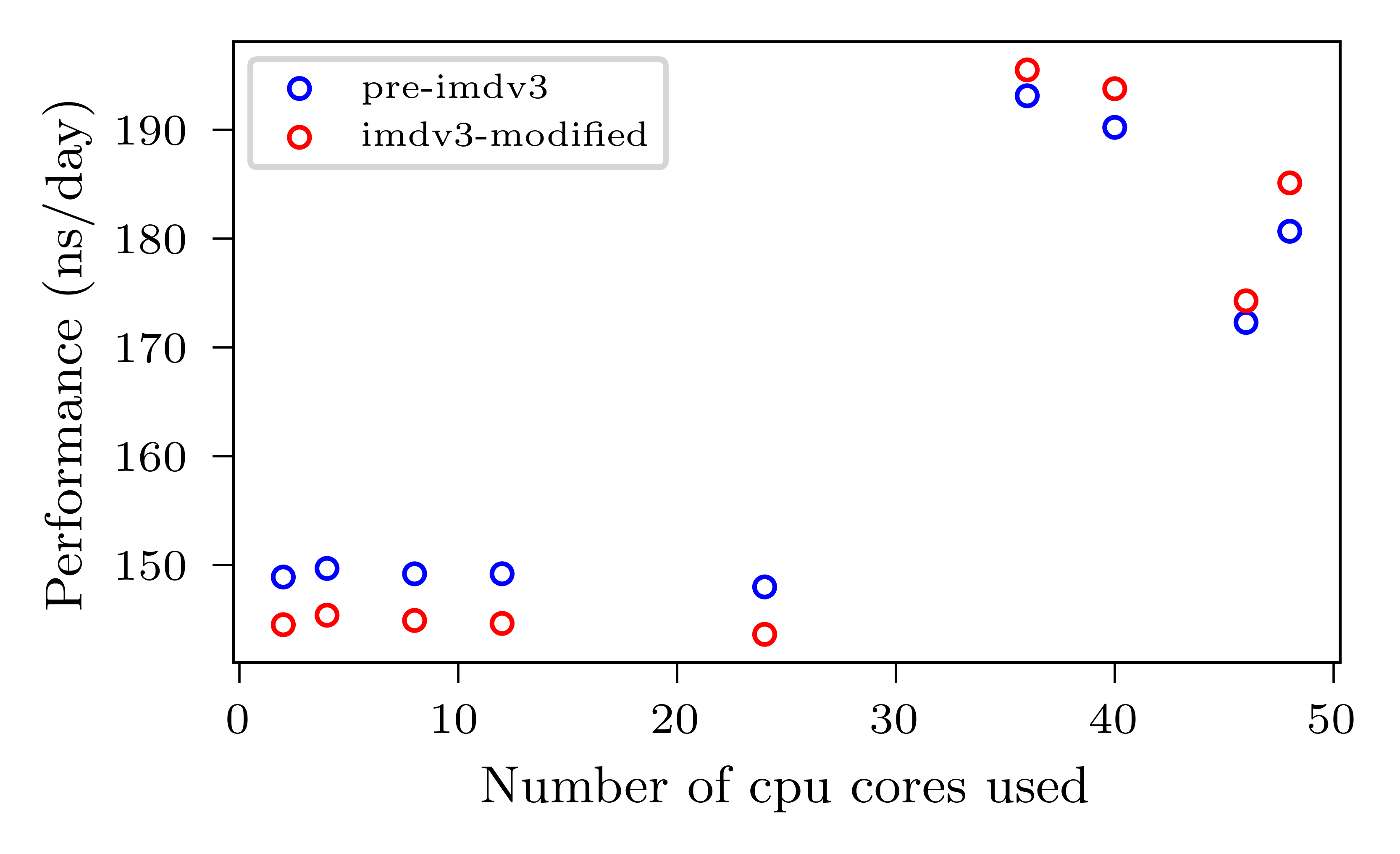}
  \caption{\textbf{Optimizing the performance of \software{NAMD}:} We run $2$ versions of \software{NAMD} viz. IMDv3-modified (`imdv3') and the pre-IMDv3 (`vanilla') version over a fixed set of computational resources. We observe that at $35$ cores (threads) under the smp-MPI mode with `GPU resident mode' enabled gives best performance. This is consistent between the $2$ versions tested, despite noted differences in the $2$ versions as discussed in the main text and above.}
  \label{figSI:NAMD-optimization}
\end{figure}

\subsection{Performance benchmark modeling}

The main text discusses the exact modeling approach (see \Cref{eq:master-sim-model}) used to capture the performance trends observed for streaming and file I/O across the $3$ simulation engines. 
We employed a least squares fitting approach to obtain the parameters for our model using the \codeinline{scipy.optimize.least_squares} \cite{scipyVirtanen2020} routine in Python.
Below, we provide the fitting parameters achieved for each simulation engine.

\begin{table}[h!]
\centering
\caption{Fit parameters for the IMDv3-benchmark performance models}
\label{tabSI:fit-params-all-engines}
\renewcommand{\arraystretch}{1.08}
\setlength{\tabcolsep}{9pt}
\begin{tabular}{@{}p{.075 \linewidth}cccccccc@{}}
	\toprule
\multirow{2}{=}{\centering Simulation engine} & \multicolumn{3}{c}{Simulation type} & \multicolumn{5}{c}{Fit parameters ($\mu s$)} \\
\cmidrule(lr){2-4}\cmidrule(lr){5-9}
& plot label & $\alpha$ (data type) & $\beta$ (I/O format) & $t_{\text{MD}}$ & $t^{(\alpha)}_\text{s}$ & $t^{(0)}_\text{s}$ & $t^{(\alpha, \beta)}_\text{IO}$ & $t^{(\alpha, \beta, 0)}_\text{IO}$ \\
\midrule
\multirow{5}{=}{\centering \software{GROMACS}}
& IMDv3-x streaming & x & - & 280 & 10 & 10 & - & - \\
& IMDv3-xvf streaming & xvf & - & 280 & 470 & 10 & - & - \\
& xtc-x file I/O & x & xtc & 280 & - & - & 1370 & 20 \\
& trr-x file I/O & x & trr & 280 & - & - & 3700 & 70 \\
& trr-xvf file I/O & xvf & trr & 280 & - & - & 7400 & 90 \\
\addlinespace
\multirow{3}{=}{\centering \software{LAMMPS}}
& IMDv3-x streaming & x & - & 602 & 1320 & 69 & - & - \\
& IMDv3-xvf streaming & xvf & - & 602 & 3980 & 69 & - & - \\
& lammpsdump-x file I/O & x & lammpsdump & 602 & - & - & 21300 & 32 \\
\addlinespace
\multirow{4}{=}{\centering \software{NAMD}}
& IMDv3-x streaming & x & - & 990 & 2570 & 20 & - & - \\
& IMDv3-xvf streaming & xvf & - & 990 & 8090 & 20 & - & - \\
& dcd-x file I/O & x & dcd & 990 & - & - & 3800 & 0 \\
& dcd-xvf file I/O & xvf & dcd & 990 & - & - & 11450 & 0 \\
\bottomrule
\end{tabular}
\end{table}

Here, fitting parameters' associated uncertainities (errors) are rounded to $1$ signifcant digit and correspondingly rounded fit values \cite{RoundingTaylor2022} are reported above.

\subsection{IMDv3 applications}
\label{subsecSI:imdv3-applications}

As discussed previously in the main text, IMDv3 based streaming can be used for various applications such as live, intermittent monitoring, {\em in-situ} analysis, and high-throughput calculation of properties and detecting rare events. Below, we describe the exact properties calculated in \Cref{fig:imd-applications} briefly.

\subsubsection{End-to-end distance of polymer}
\label{subsubsubsecSI:end-end-polymer}

In this example (see \Cref{fig:imd-applicationsa}), polymer ID-number $1$ (randomly chosen for the example) is monitored by calculating its end-to-end distance with appropriate periodic boundary conditions (PBC) based corrections. The exact calculation can be found in the \package{IMDv3-applications} repository.

\begin{equation}
   l_{\text{end-end}} = {\left\lVert \vect{r}_1 - \vect{r}_N \right\rVert}_\text{pbc}
\label{eq:end-end-distance}
\end{equation}

where $\vect{r}_1$ and $\vect{r}_N$ are the position vectors corresponding to the first and last monomer in the polymer, respectively.

\subsubsection{VACF of water}
\label{subsubsubsecSI:vacf-water}

Here (see \Cref{fig:imd-applicationsb}), the average velocity autocorrelation function (VACF) of a single water molecule in the system is calculated over a fixed lag time period and visualized at various simulation times. VACF here is defined as,

\begin{equation}
   \mathrm{VACF}_i(\tau) = \frac{\left\langle \vect{v}_i(t) \cdot \vect{v}_i(t+\tau) \right\rangle_t}{\left\langle \vect{v}_i(t) \cdot \vect{v}_i(t) \right\rangle_t}
\label{eq:vacf}
\end{equation}

where $\vect{v}_i(t)$ is the velocity of the $i^{th}$ water molecule at time $t$, and $\langle \cdot \rangle_t$ denotes an average over all time points $t$ in the simulation so far. We compare $\mathrm{VACF}_i$ with the time and system averaged version achieved by averaging over all water molecules in the system as shown in \Cref{fig:imd-applicationsb}.

\subsubsection{Ion Current through membrane}
\label{subsubsubsecSI:ion-current-membrane}

Here (see \Cref{fig:imd-applicationsc}), we calculate and visualize the ion current (nA) across the membrane pore, defined as follows:

\begin{equation}
   \mathrm{I}_i(t) = \frac{1}{L_z \Delta t} \sum_{j=1}^{N_i} q_j \left( z_j(t+\Delta t) - z_j(t) \right)_\text{pbc}
\label{eq:ion-current}
\end{equation}

where $I_i(t)$ is the ion current from a certain ion group like cations, anions or all ions. These ion currents are calculated across a time interval $\Delta t$ in the $z$-direction, with $L_z$ being the length of the simulation box in $z$. The summation sums a product of charge and PBC-aware displacement terms across all different ions and ion types belonging to group $i$.

\section{Software availability and versions used}
\label{secSI:software-avail}

Below, we have listed the particular software and package versions for our simulation and client implementations of IMDv3 alongside Zenodo archive links.
We also list links for benchmarking data, fitting analysis and IMDv3 example applications shown in \Cref{fig:imd-applications}

\begin{table*}[t]
\centering
\caption{Software availability and repositories for IMDv3 enabled simulation engines, client packages and benchmarking tools}
\label{tabSI:software-availability}
\begin{tabular}{>{\centering\arraybackslash}p{0.15\linewidth}>{\centering\arraybackslash}p{0.1\linewidth}>{\centering\arraybackslash}p{0.25\linewidth}p{0.5\linewidth}}
\toprule
\multirow{2}{*}{Category} & \multirow{2}{=}{Software/ package} & \multirow{2}{*}{Resource type} & \multicolumn{1}{c}{\multirow{2}{*}{Weblink}} \\
& & & \\
\midrule
\multirow[c]{14}{=}{\centering Simulation engine} & \multirow{5}{*}{\vspace{-3em}\software{GROMACS}} & Source code & \url{https://gitlab.com/gromacs/gromacs} \\[3pt]
\cmidrule(l){3-4}
& & pre-IMDv3 source commit & \url{https://gitlab.com/gromacs/gromacs/-/tree/55f3a10f} \\
& & pre-IMDv3 branch (`vanilla') & \url{https://gitlab.com/heydenlabasu/streaming-md/gromacs/-/tree/vanilla-benchmarking?ref_type=heads} \\[3pt]
\cmidrule(l){3-4}
& & latest IMDv3 branch (`imdv3')& \url{https://gitlab.com/heydenlabasu/streaming-md/gromacs/-/tree/imdv3-benchmarking?ref_type=heads} \\
& & Zenodo archive & \url{https://doi.org/10.5281/zenodo.20128097} \\
\cmidrule(l){2-4}
& \multirow{5}{*}{\vspace{-2em}\software{LAMMPS}} & Source code & \url{https://github.com/lammps/lammps} \\[3pt]
\cmidrule(l){3-4}
& & pre-IMDv3 source commit & \url{https://github.com/Becksteinlab/lammps/tree/528770f} \\
& & pre-IMDv3 branch (`vanilla') & \url{https://github.com/Becksteinlab/lammps/tree/vanilla-benchmarking} \\[3pt]
\cmidrule(l){3-4}
& & latest IMDv3 branch (`imdv3')& \url{https://github.com/Becksteinlab/lammps/tree/imdv3-benchmarking} \\
& & Zenodo archive & \url{https://doi.org/10.5281/zenodo.20128193} \\
\cmidrule(l){2-4}
& \multirow{4}{*}{\vspace{-2em}\software{NAMD}} & Source code & \url{https://gitlab.com/tcbgUIUC/namd} \\[3pt]
\cmidrule(l){3-4}
& & pre-IMDv3 source commit & \url{https://gitlab.com/tcbgUIUC/namd/-/tree/11b2bdcc} \\
& & pre-IMDv3 branch (`vanilla')& \url{https://gitlab.com/tcbgUIUC/namd/-/tree/vanilla-benchmarking?ref_type=heads} \\[3pt]
\cmidrule(l){3-4}
& & latest IMDv3 branch (`imdv3')& \url{https://gitlab.com/tcbgUIUC/namd/-/tree/imdv3-benchmarking?ref_type=heads} \\
\midrule
\multirow{6}{*}{Client} & \multirow{3}{*}{\vspace{-1em}\package{imdclient}} & Source code & \url{https://github.com/Becksteinlab/imdclient} \\[3pt]
\cmidrule(l){3-4}
& & latest IMDv3 branch & \url{https://github.com/Becksteinlab/imdclient/tree/v0.2.3} \\[3pt]
& & Zenodo archive & \url{https://doi.org/10.5281/zenodo.20275368} \\
\cmidrule(l){2-4}
& \multirow{3}{*}{\vspace{-1.5em}\package{MDAnalysis}} & Source code & \url{https://github.com/MDAnalysis/mdanalysis} \\[3pt]
\cmidrule(l){3-4}
& & latest IMDv3 branch & \url{https://github.com/MDAnalysis/mdanalysis/tree/refs/tags/package-2.10.0} \\[3pt]
& & Zenodo archive & \url{https://doi.org/10.5281/zenodo.17382469} \\
\midrule
\multirow{4}{=}{IMDv3 benchmarking/ applications} 
& \multirow{2}{=}{IMDv3-performance-tests} & Source code & \url{https://github.com/HeydenLabASU-collab/IMDv3-performance-tests} \\[3pt]
\cmidrule(l){3-4}
& & Zenodo archive & \url{https://doi.org/10.5281/zenodo.20090659} \\
\cmidrule(l){2-4}
& \multirow{2}{=}{IMDv3-applications} & Source code & \url{https://github.com/HeydenLabASU-collab/IMDv3-applications} \\[3pt]
\cmidrule(l){3-4}
& & Zenodo archive & \url{https://doi.org/10.5281/zenodo.20127933} \\
\bottomrule
\end{tabular}
\end{table*}

\end{document}